\documentclass[12pt,letterpaper]{article}
\usepackage{graphicx,array}
\usepackage{url}
\usepackage{color}
\usepackage{latexsym}
\usepackage{amsthm}
\usepackage{amsmath}
\usepackage{amssymb}
\usepackage{amsfonts}
\usepackage[numbers,sort&compress]{natbib}
\usepackage{bm}
\usepackage{slashed}
\usepackage{mathrsfs}
\usepackage{enumerate}
\usepackage{tikz}
\usepackage{siunitx}
\usepackage{mdframed}
\usepackage{setspace}  
\usepackage{esvect}

\usepackage[utf8x]{inputenc}%
\usepackage{tcolorbox}%

\usepackage{hyperref} %Automatically links \label and \ref commands; Always load last
\hypersetup{
    colorlinks=true,       % false: boxed links; true: colored links
    linkcolor=red,          % color of internal links
    citecolor=blue,        % color of links to bibliography
    filecolor=magenta,      % color of file links
    urlcolor=blue           % color of external links
}
\usepackage[all]{hypcap} %Link navagates to top of figure instead of caption (below fig)

\usepackage{natbib}
\newcommand{\nc}{\newcommand}

\nc{\beq}{\begin{equation}}  
\nc{\eeq}{\end{equation}}  
\nc{\beqa}{\begin{eqnarray}}  
\nc{\eeqa}{\end{eqnarray}}  
\nc{\bit}{\begin{itemize}}  
\nc{\eit}{\end{itemize}}  

\def\GeV{\mathrm{GeV}}     % GeV
\newcommand{\eg}{{\it e.g.}}
\newcommand{\ie}{{\it i.e.}}
\newcommand{\Mpl}{M_{\rm pl}}

\usepackage{floatrow}
\newfloatcommand{capbtabbox}{table}[][\FBwidth]

\usepackage{blindtext}

\newcommand{\lc}{\lambda_{\phi S}}
\newcommand{\lbc}{\overline{\lambda}_{\phi S}}
\newcommand{\lphi}{\lambda_\phi}
\newcommand{\ls}{\lambda_S}
\newcommand{\ms}{m_{S,0}}
\newcommand{\mus}{\mu_0}

\newcommand{\rb}{\overline{r}}
\newcommand{\rbc}{\rb_b}

\title{ 
 {\bf Q-Monopole-Ball:} \\
 {\bf \large A Topological and Nontopological Soliton}
\author{\large Yang Bai$^{\,\star}$, Sida Lu$\,^\diamond$, and Nicholas Orlofsky$^{\,\dagger}$}
\date{\small \it 
$^\star$Department of Physics, University of Wisconsin-Madison, Madison, WI 53706, USA\\
$^\diamond$Department of Physics, Tel Aviv University, Tel-Aviv 69978, Israel \\
$^\dagger$Department of Physics, Carleton University, Ottawa, ON K1S 5B6, Canada \\
}
}

\begin{document}

\maketitle

\setlength{\parskip}{0.2ex}

\begin{abstract}	
Magnetic monopoles and Q-balls are examples of topological and nontopological solitons, respectively. A new soliton state with both topological and nontopological charges is shown to also exist, given a monopole sector with a portal coupling to an additional scalar field $S$ with a global $U(1)$ symmetry. This new state, the Q-monopole-ball, is more stable than an isolated Q-ball made of only $S$ particles, and it could be stable against fissioning into monopoles and free $S$ particles. Stable Q-monopole-balls can contain large magnetic charges, providing a novel nongravitational mechanism for binding like-charged monopoles together. They could be produced from a phase transition in the early universe and account for all dark matter.
\end{abstract}

\thispagestyle{empty}  
\newpage  
  
\setcounter{page}{1}  
%
%\begingroup
%\hypersetup{linkcolor=black,linktocpage}
%\tableofcontents
%\endgroup
%
%\newpage

%==================================
% Introduction
%==================================
\section{Introduction}\label{sec:Introduction}

Quantum field theory offers a rich variety of possibilities for still-undiscovered soliton states beyond the Standard Model.
Many of these states can be cosmologically long-lived relics if formed in the early universe.
Among these are \textit{nontopological} solitons such as Q-balls.
For a scalar field with nonlinear interactions and charged under a global symmetry, it has long been pointed out~\cite{Rosen:1968mfz,Friedberg:1976me,Coleman:1985ki} that there may exist a classical solution where the scalar field condenses and forms macroscopic states with large global charge (see~\cite{Lee:1991ax,Nugaev:2019vru} for reviews). 
Such configurations are energetically preferred against decaying into free particles, and hence their stabilities are ensured by global charge conservation. 
Examples of such configurations exists, for example, in the Minimal Supersymmetric Standard Model (MSSM) where the squarks or sleptons have nonlinear interactions~\cite{Kusenko:1997si} and carry baryon or lepton number. 
Also, in a Higgs-portal dark matter scenario where the dark sector contains a complex scalar field, dark matter may live as solitons in which the electroweak symmetry is restored~\cite{Ponton:2019hux}. 

Another possible solitonic state is the magnetic monopole, which carries \textit{topological} charge. 
After being proposed by Dirac~\cite{Dirac:1931kp} to explain the observed quantization of electric charge, monopoles have drawn the attentions of physicists from both theoretical and experimental directions. 
On the theory side, the simplest model of monopoles was realized by 't Hooft and Polyakov~\cite{tHooft:1974kcl, Polyakov:1974ek} in a spontaneously broken $SU(2)$ gauge theory, and has been applied to the Grand Unification Theory~\cite{Georgi:1974sy}. In electroweak theory, a different monopole topology has also been pointed out~\cite{Cho:1996qd}. 
On the experimental side, monopoles have been searched for in quantum interference devices~\cite{Cabrera:1982gz} through detecting the quantized jump of magnetic flux, in detectors or materials that record the tracks of monopoles \cite{MACRO:2002jdv,Ghosh:1990ki,Aartsen:2015exf}, and at the Large Hadron Collider~\cite{MoEDAL:2014ttp,Gould:2017zwi} through direct production from ion collisions. For a recent review, see \cite{Mavromatos:2020gwk}.

In this work we propose a novel soliton state, the Q-monopole-ball (QMB), which inherits the features of both the two aforementioned examples. 
In a theory with an unbroken global symmetry and a spontaneously broken gauge symmetry that admits monopoles, it is possible to obtain a macroscopic state that is charged \textit{both topologically and nontopologically}. 
This is achieved by introducing a quartic scalar interaction between the scalar field of the global symmetry and the scalar field breaking the gauge symmetry.
From a bottom-up perspective, the model is fairly minimal, invoking only a gauged scalar to provide for the existence of a monopole (which is well motivated by charge quantization) and one new gauge-singlet scalar, then including a renormalizable scalar portal coupling between these two scalars, which should exist in the absence of any symmetry forbidding it.
From a top-down perspective, such a setup can be naturally realized by a symmetry-breaking pattern of $G/H$, where $G$ contains both gauge symmetries and a global $U(1)$ symmetry, and $H$ contains the unbroken $U(1)_\text{em}$ electromagnetic gauge symmetry and the global $U(1)$ symmetry.
Note, this setup differs from ``gauged Q-balls'' \cite{Lee:1988ag,Gulamov:2015fya,Brihaye:2015veu,Heeck:2021zvk}, where the global nontopological charge is gauged and anyways only carries the electric-charge analog of its gauge group.

Depending on the size of its global charge $Q$, a QMB may behave mainly like either a Q-ball or a monopole, which can be interpreted as one or more monopoles bound inside a large Q-ball or a small Q-ball bound inside a monopole, respectively. 
The existence and stability of such bound states  can be understood qualitatively. Both an isolated Q-ball and a monopole tend to have the spontaneously broken gauge symmetry restored in their interiors, which costs vacuum energy. Having the Q-ball and monopole ``overlap'' to form a bound state reduces the volume of space in the false vacuum, reducing the overall energy of the system. Additionally, for a monopole bound inside a larger-radius Q-ball, the magnetic charge is free to ``spread out'' over a larger volume, reducing the energy contained in the magnetic field.

For monopoles with magnetic charge $q=2$, analytic expressions for the spherically symmetric field configurations only exist in the Bogomol'nyi-Prasad-Sommerfield (BPS) limit~\cite{Bogomolny:1975de, Prasad:1975kr}, when the self interaction of the gauged scalar field vanishes. Away from this limit, spherical solutions can be readily obtained numerically.
In this work, we apply a similar numerical treatment to QMBs, which are also spherically symmetric when $q=2$. 
We show that QMBs with $q=2$ are always more stable than Q-balls of the same global charge $Q$. Additionally, smaller-$Q$ solutions exist for QMBs than for isolated Q-balls.

It is known that a monopole can have multiple magnetic charges, {\it i.e.} $q>2$, although such monopole configurations cannot be spherically symmetric~\cite{Weinberg:1976eq}.
It has been proposed that a large number of BPS monopoles may cluster into a spherical bag structure~\cite{Bolognesi:2005rk}, where the monopoles reside on a quasi-regular lattice on the spherical surface~\cite{Lee:2008ze} (see also \cite{Manton:2011vm,Evslin:2011tr,Evslin:2012fe}). Such states exist only in the BPS limit, where repulsive magnetic forces and attractive Yukawa forces (mediated by the massless symmetry-breaking scalar field) between monopoles cancel. However, gravitational forces could stabilize bound states of non-BPS monopoles, allowing for gravitational magnetic bag and magnetic black hole states \cite{Bolognesi:2010xt,Lee:1994sk,Maldacena:2020skw}. Nambu's dumbbell construction \cite{Nambu:1977ag} and cosmic necklaces \cite{Hindmarsh:1985xc,Berezinsky:1997td,Hindmarsh:2016dha} offer other bound state alternatives where non-BPS monopoles can be bound by string(s), although their geometry differs from the compact, approximately spherical configuration of interest here.
	
Our work proposes a novel nongravitational way to stabilize large-magnetic-charge configurations away from the BPS limit by having them bound in QMBs,
without attempting to decipher the nonspherical field configurations of QMBs with $q>2$. 
As long as the magnetic field energy is not too large, it can be energetically favorable for many monopoles to be bound in a larger Q-ball, according to the energetic arguments presented above.
The typical size of the magnetic charge in a QMB, on the other hand, relies on the details of the QMB formation and evolution, which we briefly discuss when examining the relic abundance of QMBs.

The organization of this paper is as follows. We first focus on QMBs with unit magnetic charge $q=2$ in Sec.~\ref{sec:unitcharge}. 
We work out the equations of motion for a QMB and the expressions for its mass $M$ and global charge $Q$, discuss its properties in both the large and small $Q$ limit, and present the computed parameter space where QMBs are stable.
Then, in Sec.~\ref{sec:large-q}, we discuss QMBs with $q>2$. 
We provide an argument why it could be energetically preferred for a QMB to have more than unit magnetic charge, and provide the corresponding expressions for QMBs' masses and global charges at $q>2$. 
Based on this, we presented the parameter region where the QMBs are cosmologically stable. 
In Sec.~\ref{sec:formation} we briefly discuss the formation of QMBs from cosmic phase transitions and the allowed parameter space for QMBs to make up all the cosmic dark matter. 
Finally in Sec.~\ref{sec:pheno} we discuss the phenomenological consequences and the detection of QMBs, before concluding in Sec.~\ref{sec:conclusion}.

\section{QMBs with unit magnetic charge ($q=2$)}
\label{sec:unitcharge}

A QMB is charged under both the $U(1)_\text{em}$ electromagnetic gauge symmetry and a $U(1)_S$ global symmetry. One could treat it as a composite state that is made of an isolated Q-ball and an isolated magnetic monopole. In our study, we use a simple toy model containing a magnetic monopole, but the qualitative results can be applied to a realistic model containing a magnetic monopole.

Consider a theory with a complex gauge-singlet scalar $S$ and a gauged $SU(2)$ triplet scalar $\phi^a$ with $a=1,2,3$ the gauge index. The most general renormalizable Lagrangian invariant under the $SU(2)$ gauge symmetry and $U(1)_S$ global symmetry is
\beqa
\label{eq:basicLag}
\mathcal{L} &=& |\partial_\mu S|^2 + \frac{1}{2} (D_\mu \phi^a)^2 - \frac{1}{4} F^a_{\mu\nu} F^{a\mu\nu} - V(S,\phi) \, ,
\\
V(S, \phi) &=& \frac{1}{8} \lphi (\phi^a \phi^a - v^2)^2 +  \frac{1}{2} \lc |S|^2 (\phi^a \phi^a) + \ls |S|^4 + \ms^2 |S|^2 \, ,
\label{eq:VSphi}
\eeqa
where $D_\mu \phi^a = \partial_\mu \phi^a + e \epsilon^{abc} A_\mu^b \phi^c$, $F_{\mu\nu}^a = \partial_\mu A_\nu^a - \partial_\nu A_\mu^a + e \epsilon^{abc} A_\mu^b A_\nu^c$, $A_\mu^a$ is the $SU(2)$ gauge field, and $e$ is the $SU(2)$ gauge coupling. Here, we take all model parameters $\lphi,\lc,\ls,\ms^2,v^2 \geq 0$, so that the vacuum expectation values (VEVs) for the scalar fields are $\langle \phi^a \phi^a \rangle = v^2$ and $\langle |S|^2 \rangle=0$.~\footnote{A somewhat similar theory containing a gauged spontaneously broken $U(1)$ and a global $U(1)$ was considered in \cite{Peter:1992dw,Battye:2021sji}. That theory results in cosmic strings rather than monopoles. In those works, the equivalent Lagrangian parameters are not all positive, so the global $U(1)$ symmetry is spontaneously broken inside the string. One could also explore the similar parameter space in the Lagrangian considered here.}

This theory admits several stable nonvacuum field configurations. One, the 't Hooft-Polyakov monopole~\cite{tHooft:1974kcl,Polyakov:1974ek}, is a topological field configuration resulting from the residual $U(1)_\text{em}$ symmetry from the spontaneous $SU(2)$ breaking (whose homotopy group satisfies $\pi_2[G/H] = \pi_2[SU(2)/U(1)]= \mathbb{Z}$). It involves only the $\phi^a$ and $A_\mu^a$ fields, with $|S|^2=0$ being irrelevant. Indeed, $\lc$ could be zero, or equivalently $S$ need not be present in the theory for this solution. Alternatively, for different sign choices in the Lagrangian parameters, the $S$ VEV could be modified inside the monopole~\cite{Bai:2020ttp}. 

Another configuration, the nontopological soliton~\cite{Friedberg:1976me} or Q-ball~\cite{Coleman:1985ki}, allows for a large global ``charge'' of $S$ owing to the global $U(1)_S$ symmetry in the theory and a nontrivial interplay with the $\phi^a$ field~\cite{Ponton:2019hux}. This soliton solution has no interplay with the gauge field $A_\mu^a$ and exists even in the limit $e=0$.

Both of the prior two solutions have been explored extensively in the literature. Here, we point out another novel solution to the classical equations of motion of fields in \eqref{eq:basicLag} and a new object: the {\it Q-Monopole-Ball}, which is charged under both the $U(1)_\text{em}$ gauge and $U(1)_S$ global symmetry. By allowing all three fields to interplay with one another, it is possible to form a mixed state containing both topological and nontopological charges labeled by $q$ and $Q$, respectively. In certain limits, this mixed state could be approximately interpreted as a monopole trapped inside a Q-ball or several $S$ quanta trapped inside a monopole. We demonstrate this solution and its various limits in the following subsections.

In this section, we focus on unit monopole charges, which admit spherical solutions. By unit charge, we mean the magnetic charge is $q_\text{mag}=2 \pi q / e$ with $q=2$. The Dirac quantization of electric and magnetic charge is satisfied for integer $q$ with $e\,q_\text{mag} = 2 \pi q$, and the spherical 't Hooft-Polyakov monopole solution corresponds to $q=2$. Higher charges $q>2$ can only be achieved with nonspherical solutions \cite{Weinberg:1976eq}, making it analytically intractable to determine their field configurations. We return to these in Sec.~\ref{sec:large-q}.

We additionally look for solutions that are stable against decaying to other states. For convenience, we label states according to their magnetic and $U(1)_S$ charges by the notation $(q,Q)$. In this notation, an ordinary monopole without $S$ field is $(2, 0)$, a free $S$ particle is (0,1), an ordinary Q-ball without gauge fields is $(0, Q)$, and a QMB can have both entries nonzero. The most important channels through which a QMB with unit magnetic charge $q=2$ can decay are
\beqa
Q\mbox{-particle decay}:&& \quad  (2, Q) \to Q \times (0,1) + (2, 0) \,,  
\label{eq:decay-to-free}\\
\mbox{Q-ball decay}:&& \quad (2, Q) \to  (0, Q) + (2, 0) \,, 
\label{eq:decay-to-Qball}\\
\mbox{1-particle decay}:&& \quad (2, Q) \to (2, Q-1) +  (0,1) \,.
\label{eq:decay-to-single-S}
\eeqa

We will see later that the first process (\ref{eq:decay-to-free}) is the most important, so we will focus on this one when discussing analytic approximations. The process (\ref{eq:decay-to-Qball}) is disfavored because Q-balls are more stable when mixed with monopoles. Additionally, Q-ball and QMB solutions generally only exist if emitting a single free $S$ particle like in process (\ref{eq:decay-to-single-S}) is energetically impossible (we show this in the following subsection).

\subsection{Classical equations of motion}

To write the equations of motion, we parametrize the fields by the following spherically symmetric ansatz depending on radius $r$:
\begin{equation}
\label{eq:field-ansatz}
\phi^a = \hat{r}^a \,v\, f(r) \, , \; \; S = e^{-i \omega t} \frac{v}{\sqrt{2}}\,s(r) \, , \; \; A_0=0 \, , \; \; A_i^a = \epsilon^{aij} \frac{\hat{r}^j}{e\, r} \, a(r) \, ,
\end{equation}
where $\hat{r}^a = r^a / r$. With a rescaling of the coordinate $\rb = v\,r$ and parameters $\Omega = \omega / v$ and $\mus=\ms/v$, the equations of motion can be written in terms of dimensionless quantities: 
\begin{align}
\label{eq:eom-gauge-field}
& a'' - \frac{1}{\rb^2} \,a (1-a) (2-a) + e^2 \,(1-a)\, f^2 = 0 \, ,
\\
& f'' + \frac{2}{\rb}\, f'  - \frac{2}{\rb^2}\, (1-a)^2\, f - \frac{1}{2}\, \lphi\, f \,(f^2-1) - \frac{1}{2}\, \lc \,s^2 f = 0 \, ,
\label{eq:EOM-f}
\\
& s'' + \frac{2}{\rb}\, s' + \Omega^2\, s - \frac{1}{2}\, \lc\, f^2 \,s - \ls\, s^3 - \mus^2 \, s = 0 \, ,
\label{eq:EOM-g}
\end{align}
where primes denote derivatives with respect to $\rb$.

For the QMB, the boundary conditions at the origin are determined by demanding that the equations of motion not diverge. At infinity, the scalar fields must take their VEVs and the gauge field must behave as a monopole. Thus, the boundary conditions are
\begin{equation}
f(0) = 0 \, , \; f(\infty) = 1 \, , \; s'(0) = 0 \, , \; s(\infty) = 0 \, , \; a(0) = 0 \, , \; a(\infty) = 1 \, .
\label{eq:bc}
\end{equation}
Notice the Neumann boundary condition for $s$ at the origin. For convenience, we define $s(0) = s_0$, which will depend on $\Omega$. These boundary conditions could be compared to those of a pure monopole, which has $s=0$ identically but the same conditions on $f$ and $a$ as in (\ref{eq:bc}). They could also be compared to those of a pure nontopological soliton, which has $a=0$ identically and a Neumann boundary condition $f'(0)=0$ replacing the Dirichlet condition on $f(0)$ in (\ref{eq:bc}).~\footnote{The pure nontopological soliton uses the gauge choice $\phi^b = \delta^{3 b} v f(r)$, which does not contain a topological charge when $a=0$, instead of the expression in (\ref{eq:field-ansatz}). So, the third term $\propto \rb^{-2}$ in (\ref{eq:EOM-f}) is absent, enabling the Neumann boundary condition without introducing a singularity in the equations of motion. Note that using the ansatz in \eqref{eq:field-ansatz} and without the gauge field, one could have a soliton state with global topological and global nontopological charges, which we do not explore here.}

The solution for $s$ can be thought of as a particle starting at rest at some large value $s_0$ at $r=0$, then rolling towards a stationary point at the origin, coming to rest there as $r \to \infty$~\cite{Coleman:1985ki}. Its effective potential is $U_\text{eff}=\Omega^2 s^2/2 - V(s,f)$, with the last term given by plugging appropriate expressions from (\ref{eq:field-ansatz}) into (\ref{eq:VSphi}). To have a nontrivial solution with $s_0 \neq 0$ that rolls monotonically to $s=0$, $s$ must have a negative mass term around $s=0$ (when $f=1$) in $U_\text{eff}$. 
Thus, $\Omega$ must satisfy
\begin{equation}
\overline{\Omega}^2 \equiv \Omega^2 - \mus^2 < \frac{\lc}{2} \, .
\label{eq:Omega-upper}
\end{equation}
Additionally, there is a minimum imposed on $|\Omega|$ by requiring $U_\text{eff}>0$ for some value $s>0$. If $|\Omega|$ is too small, the quartic $\lphi$ term dominates $U_\text{eff}$ and makes it negative, such that there is no  Q-ball solution for which the field rolls towards the origin.~\footnote{See the left panel of Fig.~4 in Ref.~\cite{Ponton:2019hux}.} This condition is derived by setting $f=0$ and finding a positive solution $s_\text{max}$ for $\partial U_\text{eff}/\partial s = 0$, then demanding $U_\text{eff} (s=s_\text{max})>0$. The result is
\begin{equation}
\overline{\Omega}^4 > \overline{\Omega}_c^4 \equiv \frac{1}{2}\, \ls\, \lphi \, . 
\label{eq:Omega-lower}
\end{equation}
As $|\overline{\Omega}|$ approaches $|\overline{\Omega}_c|$ from above, the $U(1)_S$ charge will increase.  One can think of the $s$ field taking a long time to roll to the origin as $\overline{\Omega}$ saturates this limit, resulting in a larger radius and charge for the Q-ball. Note that the combination of (\ref{eq:Omega-upper}) and (\ref{eq:Omega-lower}) implies
\begin{equation}
\lc > \sqrt{2\,\ls\, \lphi} \, .
\label{eq:lc-min}
\end{equation}
Otherwise, no Q-ball or QMB solution exists.

Once a solution is determined, the global $U(1)_S$ charge $Q$ of the solution is
\begin{equation}
Q = i \int d^3 x \, \left(S^\dagger\, \partial_t\, S - S\, \partial_t \,S^\dagger \right) = 4 \pi \,\Omega \int_{0}^{\infty} d\rb \, \rb^2 s^2 \, .
\label{eq:charge}
\end{equation}
Classically, any irrational value of $Q$ is allowed by the equations of motion. However, because the $S$ particles are quantized, $Q$ can only take on integer values in practice. For clarity, we will limit ourselves to $\Omega>0$ in future discussions. A sign flip leads to otherwise identical solutions with opposite $Q$. The solution's mass is
\begin{equation}
\begin{aligned}
M = 4 \pi v \int_{0}^{\infty} d\rb \, \Bigg{\{} & \frac{1}{2 e^2} \left[ 2 a'^2 + \frac{1}{\rb^2} \left( 2a - a^2 \right)^2 \right] +  \left(1-a\right)^2 f^2 + \rb^2 \left[ \frac{1}{2} f'^2 + \frac{1}{8} \lphi (f^2-1)^2 \right]
\\
& + \rb^2 \left[\frac{1}{2} s'^2 + \frac{1}{2} \Omega^2 s^2 + \frac{1}{4} \lc f^2 s^2 + \frac{1}{4} \ls s^4 + \frac{1}{2} \mus^2 s^2 \right] \Bigg{\}} \, .
\label{eq:mass}
\end{aligned}
\end{equation}

That the 1-particle decay channel (\ref{eq:decay-to-single-S}) is stable can be understood from these equations. Note that $\partial M/\partial \Omega = \Omega v (\partial Q / \partial \Omega)$. The change in the QMB energy for a change in its charge $\Delta Q=1$ is thus $\Delta M \approx \Delta Q (\partial M/\partial Q) = \Omega v < \sqrt{\ms^2+\lc v^2/2}$, using (\ref{eq:Omega-upper}) in the last inequality. Thus, there is more rest mass energy in a free particle than energy contained in the bound particle. This argument is strictly true in the classical theory where $\Delta Q \to 0$ is allowed, although it need not necessarily hold for quantized $\Delta Q = 1$ (numerically, we found no counterexamples in our scans). Indeed, this does not guarantee stability in the $Q$-particle decay channel (\ref{eq:decay-to-free}), where $\Delta Q$ is too large for the approximation of constant partial derivative to hold.

Let us briefly comment on the bare Lagrangian mass $\mus$. The solutions for the field profiles $a(\rb),f(\rb),s(\rb)$ for a fixed value of $\overline{\Omega}$ with $\mus>0$ are identical to the field profile solutions when $\mus=0$ and $\Omega = \overline{\Omega}$. However, the mass and charge for these two cases differ. Defining $M_0(\overline{\Omega}) \equiv M|_{\mus=0, \, \Omega=\overline{\Omega}}$ in (\ref{eq:mass}), the mass for arbitrary $\mus$ can be written in terms of $M_0$ and $Q$ as
\begin{equation}
M = M_0(\overline{\Omega}_Q) + \frac{\mus^2}{\sqrt{\overline{\Omega}_Q^2 + \mus^2}} Q \,v \, .
\label{eq:mass-mS0}
\end{equation}
This will have an effect on the stability of the QMB (or Q-ball). Considering the most relevant decay channel (\ref{eq:decay-to-free}), the binding energies for the cases $\mus =0$ and $\mus > 0$ are
\beqa
\Delta M = \left\{ \begin{array}{l r} M_0(\overline{\Omega}_Q) - \sqrt{\frac{\lc}{2}} \,Q \,v - M_{(2,0)} \, , \, & \mus = 0  
\vspace{3mm}
\\ 
M_0(\overline{\Omega}_Q) - \left(\sqrt{\frac{\lc}{2}+\mus^2} - \frac{\mus^2}{\sqrt{\overline{\Omega}_Q^2+\mus^2}} \right) \,Q \,v - M_{(2,0)} \, , \, & \mus > 0 \end{array} \right. \, .
\eeqa
Here, $M_{(2,0)}$ is the isolated monopole mass. Thus, for the purposes of binding energy and stability in this channel, the results when $\mus=0$ can be translated to the results when $\mus >0$ by replacing $\lc$ with $\lbc$ satisfying
\begin{equation}
\sqrt{\frac{\lbc}{2}} = \sqrt{\frac{\lc}{2}+\mus^2} - \frac{\mus^2}{\sqrt{\overline{\Omega}_Q^2+\mus^2}} \, .
\label{eq:lbc}
\end{equation}
It is easy to show using (\ref{eq:Omega-upper}) that $\lc \geq \lbc > 0$, where the first inequality is only equal when $\mus=0$, and that $\lbc$ is a monotonically decreasing function of $\mus$. In short, increasing $\mus$ leads to decreased binding energy, making QMBs (or Q-balls) less stable. This can increase the minimum stable charge satisfying $\Delta M>0$. However, if a QMB is stable for any $Q>0$ when $\mus=0$, a similar QMB with $\mus>0$ and all other Lagrangian parameters fixed should also be stable. This is because the extra mass contribution from $\mus$ in (\ref{eq:mass-mS0}) goes to zero as $Q \to 0$. For the remainder of this paper, we will take $\mus=0$ for convenience. The mass, charge, and stability for otherwise identical QMBs with $\mus > 0$ can be determined using these relationships.

An interesting relation for the mass can be derived for the $s$-dependent terms in $M$. By substituting the equation of motion (\ref{eq:EOM-g}) to eliminate $\lc$ then integrating by parts, the $s$-dependent part of the mass [all terms in the second line of (\ref{eq:mass})] can be written as
\begin{equation}
M_S = Q\, \Omega\, v - 4 \pi v \int_{0}^{\infty} d\rb \, \rb^2 \frac{1}{4} \,\ls\, s^4 \, .
\label{eq:MS-eom}
\end{equation}
Note that $\Omega$ depends on $Q$ and $s(\rb)$ via (\ref{eq:charge}). Nevertheless, this equation gives the qualitatively correct behavior for a mass of a Q-ball with large charge: $M \propto Q v$ [note that the $\ls$ term in (\ref{eq:MS-eom}) is cancelled by the $\lphi$ term in the first line of (\ref{eq:mass}), at least in the step-function approximation in Sec.~\ref{sec:large-Q} using (\ref{eq:stepfcn}) and (\ref{eq:g0-largeQ})]. Indeed, we will see in the following subsection that as $Q$ increases, $\Omega$ saturates to its lower limit (\ref{eq:Omega-lower}) and can be treated as approximately constant.

\subsection{Large Q-ball charge}
\label{sec:large-Q}

It is useful to obtain analytic estimates for how the mass and charge depend on the parameters in the equations of motion. One estimate is obtained in the large $Q$ limit by assuming the fields $f(\rb)$ and $s(\rb)$ behave like step functions and the $a(\rb)$ field behaves like a power-law function:
\beqa
\label{eq:stepfcn}
f(\rb) \approx  \left\{ \begin{array}{l r} 0 \, , \, & \rb<\rbc \\ 1 \, , \, & \rb>\rbc \end{array} \right. \, , \;\; \;
s(\rb) \approx \left\{ \begin{array}{l r} s_0 \, , \, & \rb<\rbc \\ 0 \, , \, & \rb>\rbc \end{array} \right. \,, \;\; \;
a(\rb) \approx  \left\{ \begin{array}{l r} \rb^2/\rbc^2 \, , \, & \rb<\rbc \\ 1 \, , \, & \rb>\rbc \end{array} \right. \,.
\eeqa
Here, the $\rb^2$ power for $a(\rb)$ is derived from \eqref{eq:eom-gauge-field} for $f(\rb)=0$ and in the limit of $a(\rb) \ll 1$. 
Then, the $S$ charge is
\begin{equation}
\label{eq:Q_stepfcn}
Q \approx \frac{4\pi}{3}\, \rbc^3 \,\Omega\, s_0^2 \, .
\end{equation}
For simplicity, we will take $\mus=0$ in the following. Then, the mass is (ignoring the gradient energies for $f(\rb)$ and $s(\rb)$, which are only important near the boundary $\rbc$ and thus small in the large $Q$ limit) 
\beq
\label{eq:M-step-contributions}
M_{(2,Q)} \approx \frac{304\,\pi\,v}{35\,e^2\,\rbc} + \frac{4 \pi}{3} \rbc^3 v \left(\frac{1}{4} \ls s_0^4 + \frac{1}{8} \lphi + \frac{1}{2} \Omega^2 s_0^2 \right)  \,.
\eeq
The first term of (\ref{eq:M-step-contributions}) can be thought of as the energy contained in the magnetic field $B$ generated by the bound monopoles, or roughly $\int d^3 x B^2/2$. The other terms are the volume contributions to the vacuum energy from $S$ and $\phi^a$ differing from their VEVs. After substituting (\ref{eq:Q_stepfcn}) to eliminate $\Omega$ in favor of $Q$ and $s_0$, the expression for $M$ is minimized (taking $\partial M/\partial s_0=\partial M/\partial \rbc=0$) by
\beqa
s_0 &\approx& \frac{1}{\rbc} \left(\frac{9\,Q^2}{16\,\pi^2\, \ls} \right)^{1/6} \, ,
\\
\rbc &\approx& \left(\frac{2432\, \pi + 105 \,(6/\pi)^{1/3}\,\ls^{1/3} \,e^2 \,Q^{4/3}}{140 \, \pi\,e^2 \,\lphi} \right)^{1/4}  ~.
\eeqa
It is useful to go to the limit $\ls^{1/3} e^2 Q^{4/3} \gg 60$ because the step function approximation would not be expected to hold in the opposite limit anyways (for the case $\ls=0$, a different ansatz is necessary, {\it c.f.}~\cite{Ponton:2019hux}). This is equivalent to neglecting the first term in (\ref{eq:M-step-contributions}), meaning that the magnetic field energy is negligible. Then,
\beqa
\rbc &\approx& \frac{(3/\pi)^{1/3}}{2^{5/12}} \, \frac{\ls^{1/12}}{\lphi^{1/4}} Q^{1/3} \, ,
\label{eq:uc-largeQ}
\\
M_{(2,Q)} \approx M_{(0,Q)} &\approx& Q \, \Omega_c \, v  
\, ,
\label{eq:M-largeQ}
\\
s_0 &\approx& \left(\frac{\lphi}{2\,\ls}\right)^{1/4} \, ,
\label{eq:g0-largeQ}
\\
\Omega &\approx& \Omega_c \, .
\eeqa
Note, $\Omega$ saturates the bound in (\ref{eq:Omega-lower}) in the large $Q$ limit, and $M \propto \rbc^3 \propto Q$ so the QMB has a fixed energy density.

In this large $Q$ limit, the system can be thought of as a small-radius monopole trapped in a large-radius Q-ball. There always exists a sufficiently large $Q$ such that the QMB configuration is stable against decay to an isolated Q-ball and monopole. As $Q$ and $\rbc$ increase, the first term in (\ref{eq:M-step-contributions}) corresponding to the magnetic field energy becomes less relevant, and the mass approaches that of a Q-ball without a magnetic charge as in (\ref{eq:M-largeQ}). Including this magnetic field energy term in $M_{(2,Q)}$, the binding energy for a monopole and Q-ball is
\begin{equation}
\Delta M = M_{(2,0)} + M_{(0,Q)} - M_{(2,Q)} \approx \frac{4 \pi v}{e} Y - \frac{304\times 2^{5/12}\,\pi^{4/3}}{35\times 3^{1/3}}\,\frac{\lphi^{1/4}}{e^2\,\ls^{1/12} }\,Q^{-1/3}\,v \, .
\label{eq:binding-single-monopole}
\end{equation}
The first term is the mass of an isolated monopole $M_\text{(2,0)} = 4 \pi v Y / e$. Here, $Y \equiv Y(\lphi/e^2)$ is monotonic with $Y(0)=1$ corresponding to the BPS limit and $Y(\infty) \approx 1.787$. The second term $\propto Q^{-1/3}$ is the mass energy contained in the gauge field $a$ coming from the first term in (\ref{eq:M-step-contributions}). The leading mass term in (\ref{eq:M-largeQ}) is canceled by the subtraction.
Thus, the mass of the combined QMB system can always be smaller than the sum of the masses of an isolated monopole and a Q-ball for sufficiently large $Q$. 

Additionally, if $(M_{(2,Q)}-M_{(2,0)})/Q < m_S = v \sqrt{\lc/2}$, where $m_S$ is the mass of the $S$ particles in the true vacuum, such QMBs are stable against decays to free $S$ particles. Using (\ref{eq:M-largeQ}) in the limit $M_{(2,Q)} \gg M_{(2,0)}$, this stability condition is equivalent to the existence condition in (\ref{eq:lc-min}). This implies that there always exists a sufficiently large $Q$ such that QMB (and Q-ball) solutions are stable if such solutions exist.

Beyond the step-function approximation, the frequency $\Omega$ asymptotes to $\Omega_c$ in the large $Q$ limit. The radius depends on $\Omega$ as $\rbc = c_1/(\Omega - \Omega_c)$, with $c_1>0$ as a numerical coefficient \cite{Ponton:2019hux}. Ignoring the magnetic field energy, the Q-ball mass is
\beqa
\label{eq:Q-ball-mass}
M_{(2,Q)} \approx M_{(0,Q)} \approx Q\,v\,\Omega \approx  Q\,v\,\left(\Omega_c + c_2 \, Q^{-1/3}\right) ~,
\eeqa
with $c_2 > 0$ related to $c_1$ and other couplings. The subleading term $\propto Q^{2/3} \propto \rbc^{2}$ is the surface energy contribution to the mass. We confirm this behavior numerically for the QMB. The above $Q$-dependence for mass means that two Q-balls with charge $Q_1$ and $Q_2$ have a binding energy of 
\beqa
\Delta M = M_{Q_1} + M_{Q_2} - M_{Q_1+Q_2} =   c_2\, v\, \left[ Q_1^{2/3} + Q_2^{2/3}  - (Q_1+Q_2)^{2/3}\right] ~.
\eeqa
This expression is always positive, so a Q-ball or QMB is stable against fissioning to multiple Q-balls of smaller charge.

\subsection{Small Q-ball charge}
\label{sec:small-Q}

In the opposite limit of small $Q$, the $S$ field only causes a small perturbation to the monopole field configuration. Thus, it is energetically preferred for a small number of $S$ particles to be trapped in a monopole. Provided they cluster near the center of the monopole where $f\sim 0$, the $S$ particles themselves have smaller mass inside (where $m_S \sim \ms$) than outside (where $m_S = \sqrt{\ms^2+\lc v^2/2}$). 

To get a sense of the small $Q$ limit, we make the following approximations. First, take $\mus=\ls=0$ and approximate
\begin{equation}
s(\rb) \approx \left\{ \begin{array}{l l} s_0 \dfrac{\sin(\Omega\, \rb)}{\Omega\, \rb} \, , \;  & \vspace{2mm} 
 \rb<\rbc=\dfrac{\pi}{\Omega} \\ 0 \, , & \rb>\rbc \end{array} \right. \, .
\label{eq:g-sinc}
\end{equation}
This is the analytic solution for $s$ when $f=0$ in (\ref{eq:EOM-g}). We will further simplify by considering the BPS monopole \cite{Bogomolny:1975de,Prasad:1975kr} (the limit $\lphi/e^2 \to 0$), for which an analytic solution is known: $f_\text{BPS}(\rb)=\coth(e \rb) - 1/(e \rb)$. This is bounded above by $f_\text{BPS}(\rb) \leq e \rb / 3$ (to good approximation, one could replace $e$ with $e+\sqrt{2 \lphi}$ in this equation and throughout the rest of this paragraph in the non-BPS limit). We assume that the small perturbation in $f$ from the $\lc$ term is unimportant. Plugging this upper limit on $f_\text{BPS}$ and (\ref{eq:g-sinc}) into the second line of (\ref{eq:mass})---what we called $M_S$ above---then minimizing with respect to $\Omega$, 
\begin{equation}
M_S \lesssim \frac{2(4 \pi^2-6)^{1/4}}{3\sqrt{3}} Q \,e^{1/2} \,\lc^{1/4} \, v \, .
\end{equation}
The numerical prefactor is $\approx 0.93$. Comparing to the mass of $Q$ free $S$ particles $Q (\lc/2)^{1/2} v$, we see that the Q-ball should be stable unless the gauge coupling is large: $e \gtrsim \lc^{1/2}$. While we have assumed $\ls=0$, we observe {\it a posteriori} that the $\ls$ term can generally be neglected compared to the $\Omega^2$ term in $M_S$ at small $Q$: $\ls s_0^4 / (\Omega^2 s_0^2) = \ls Q /(2 \pi^2)$, which for small enough $Q$ and perturbative $\ls$ should always be less than unity.

In the regime of large gauge coupling, $e \gtrsim \lc^{1/2}$, the prior approximation that $f=f_\text{BPS}$ is not valid because the Q-ball radius is comparable to the monopole radius. The effects of the $S$ field on the $\phi$ VEV must be included. To see this, we first simplify the ansatz in (\ref{eq:g-sinc}) to make it more analytically tractable:
\begin{equation}
s(\rb) \sim \left\{ \begin{array}{l l} s_0 \, , \;  & \rb<\rbc=\dfrac{\pi}{\Omega} \vspace{2mm} \\ 
 0 \, , & \rb>\rbc \end{array} \right. \, .
\end{equation}
The scalar potential (\ref{eq:VSphi}) can be rewritten to reflect the dependence of the $\phi$ VEV on $S$:
\begin{equation}
\frac{1}{8} \lphi \left[\phi^a \phi^a - (v^2 - 2 \lc \lphi^{-1} |S|^2) \right]^2 + \frac{1}{8} \lphi \left[ v^4 - (v^2 - 2\lc \lphi^{-1} |S|^2)^2 \right] + \ms^2 |S|^2 +\ls |S|^4 \, .
\end{equation}
Defining the VEV inside the monopole for sufficiently small $s_0$ as 
\begin{equation}
\overline{v}^2 \equiv v^2 \left(1 - \lc \lphi^{-1} s_0^2 \right) \, ,
\label{eq:v-modified}
\end{equation}
and assuming the monopole and Q-ball radii are comparable, the mass is given approximately by [using (\ref{eq:Q_stepfcn})]
\begin{equation}
M_{(2,Q)} = \frac{4 \pi \overline{v}}{e} Y + \left[\frac{\pi}{2} + \left(\ls - \frac{\lc^2}{2\lphi}\right) \frac{3}{16 \pi^2} Q \right] \frac{Q v}{\rbc} + \left(\frac{\lc}{4 \pi} + \frac{\mus^2}{2 \pi} \right) Q \, v\, \rbc \, .
\end{equation}
The first term is $M_{(2,0)}$ with $v$ set to $\overline{v}$. Taking $\mus=0$ and neglecting higher powers of $Q$, the mass $M$ is minimized by $\rbc \simeq \pi \sqrt{2/\lc}$. To justify the assumption of comparable monopole and Q-ball radii so that $v$ can be replaced by $\overline{v}$, we require $\rbc \gtrsim 1/e$, or $\lc/e^2 \lesssim 2 \pi^2$. Indeed, this is exactly where the prior approximation ($f=f_\text{BPS}$) appeared to give unstable Q-balls, revealing that approximation was invalid in this limit. With this value for $\rbc$, we obtain the mass at leading order in $Q$ after a binomial expansion on $\bar{v}$:
\begin{equation}
M_{(2,Q)} \simeq \frac{4 \pi v}{e} Y + Q v \left(\sqrt{\frac{\lc}{2}} - \frac{3 \,\lc^2}{4 \pi^3\,e\, \lphi} Y \right) \, .
\end{equation}
This indicates that the mixed QMB is stable in this limit because the free monopole mass is $4 \pi v Y / e$ and the free $S$ particle mass is $\sqrt{\lc /2 }\,v$.

The above arguments cover both the cases when the Q-ball radius is either larger or smaller than the monopole radius and indicate that the QMB is stable in both cases. However, they assume that the monopole fields are relatively unchanged by the presence of the $s$ field---$f \approx f_\text{BPS}$ for the first case and $\overline{v}>0$ in (\ref{eq:v-modified}) for the second case. When these assumptions are violated, on the other hand, analytic approximation is intractable, and we must resort to qualitative explanations to explain the stability of small-$Q$ QMBs.
First, consider the BPS limit $\lphi=0$. When $S$ particles are added to the monopole starting from $Q=0$, the region where $\phi=0$ is expanded because the $\lc$ term contributes a positive mass-square to $\phi$. There is no vacuum energy cost to expanding the $\phi=0$ region in this limit. Thus, all charges $Q$ are stable in that they decrease the overall energy of the system by decreasing the mass energy of $S$ particles. This explains why the analytic approximation above where we took $f=f_\text{BPS}$ breaks down in the large $e$ limit, where the isolated monopole radius $\propto (e v)^{-1}$ is small.

Going away from the BPS limit so that $\lphi>0$, there is now a vacuum energy cost to expanding the radius of the region with $\phi=0$. As $\lphi$ increases, the radius where $f \sim 0$ (which goes as $\rb \sim \min [e^{-1},\lphi^{-1/2}]$ for an isolated monopole) shrinks, and it also costs more vacuum energy to increase its radius. Thus, at sufficiently large $\lphi$, the region with $f \sim 0$ has too narrow of a radius to admit bound states of $S$ particles. Only at sufficiently large $Q$, where there is an additional energy saving from having many $S$ particles together in a Q-ball, do bound states exist.

\subsection{Numerical results}

\begin{figure}[t]
	\centering
	\includegraphics[width=0.48\textwidth]{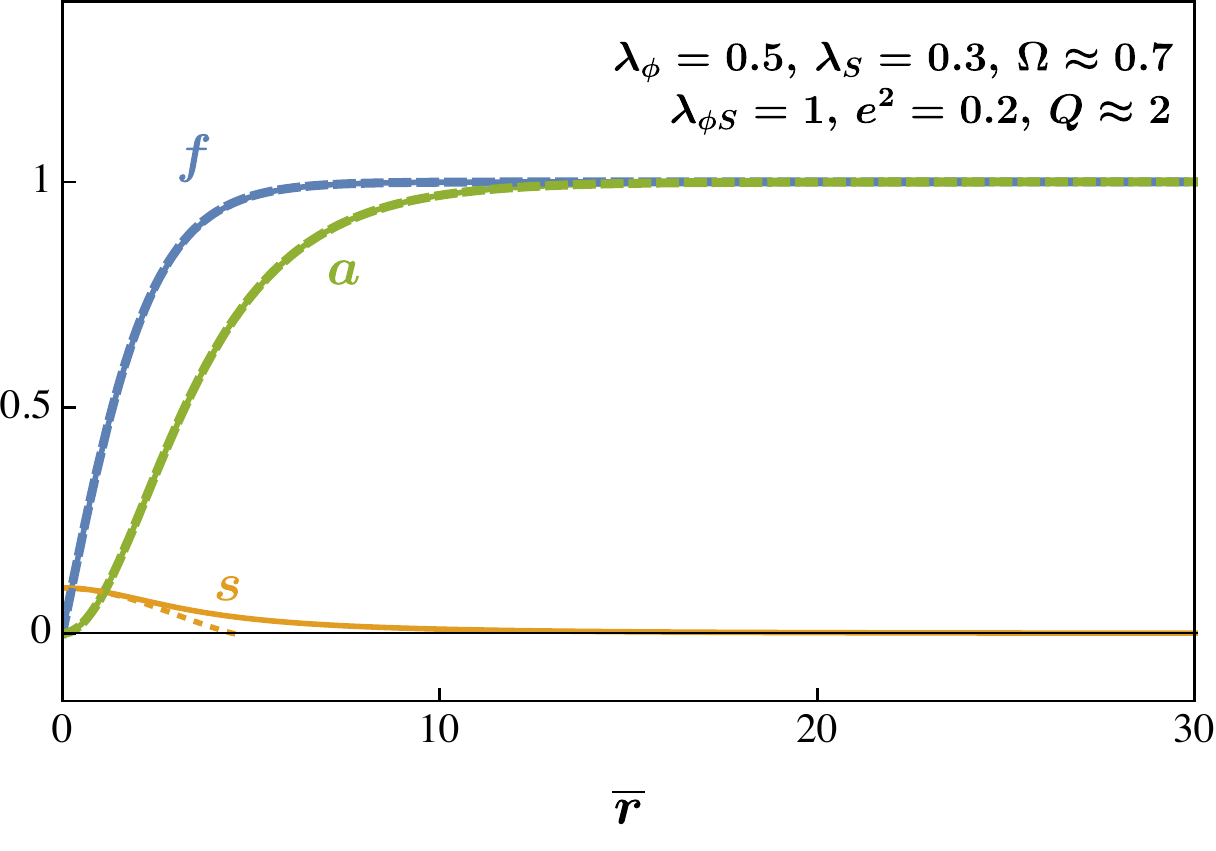} \hspace{3mm} 
	\includegraphics[width=0.48\textwidth]{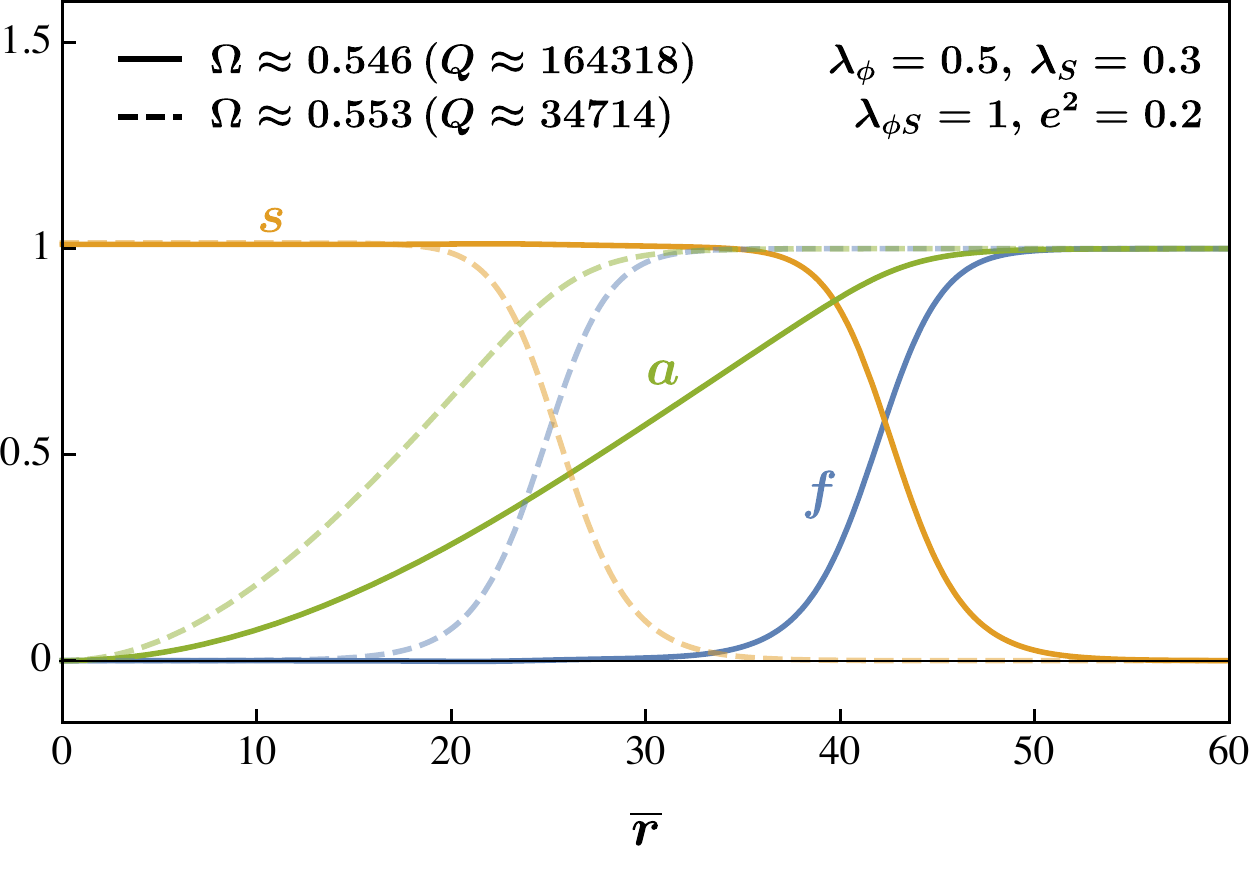}
	\caption{Example QMB field configurations for small (left) and large (right) $S$ charge $Q$. Both panels take $\ms=0$ and $q=2$. In the left panel, the dashed blue and green lines indicate the $Q=0$ solution for an isolated monopole with the same $\lphi$ and $e$, which is barely distinguishable from the $Q=2$ field profiles (solid) for $a$ and $f$. The dotted yellow line shows $s_0 \sin(\Omega \rb)/(\Omega \rb)$, which well approximates the small-radius solution for $s$ in the small $Q$ regime. In the right panel, two different choices for large Q are shown in solid and dashed lines.}
	\label{fig:field-config}
\end{figure}

To be more quantitative, we solve the system numerically using the finite difference method. We use the rescaling in \cite{Heeck:2021zvk} to recast the semi-infinite boundary at $\rb=\infty$ into a finite range, taking $y = \rb/(1+\rb/a)$ as the dependent variable with $a$ an arbitrary positive constant and $y \in [0,a]$. 
We scan for solutions over the parameters $\lc$, $\ls$, $\lphi$, $e$, and $\Omega$. When scanning along a given parameter, we take small steps in that parameter and use the previous solution as the initial guess for the next solution. This recursive approach allows us to scan to points near the limits in (\ref{eq:Omega-upper}--\ref{eq:lc-min}). 
We focus on the case $\mus=0$ in the plots presented here, but we will later comment on the changes when $\mus>0$.

Example numerical solutions are shown in Fig.~\ref{fig:field-config}, focusing on the small and large $Q$ limits. At small $Q$ (left panel), the $f$ and $a$ fields for the QMB are barely distinguishable from those of an isolated monopole when $Q=0$ (the dashed lines in the left panel). The $s$ field is well-approximated by (\ref{eq:g-sinc}), shown dotted. When $Q$ is large (right panel), the radius where $f$ and $a$ differ from unity is significantly expanded to be of comparable radius to the $s$ profile, agreeing with the approximation in (\ref{eq:stepfcn}). Increasing $Q$ in this limit causes the radius to increase but has almost no effect on $s_0$, matching the analytic estimates in (\ref{eq:uc-largeQ}) and (\ref{eq:g0-largeQ}). That $s_0 \approx 1$ in the right panel is a numerical accident---other choices of $\lphi$ and $\ls$ would result in different $s_0$ at large $Q$ according to (\ref{eq:g0-largeQ}).

\begin{figure}[t]
	\centering
	\includegraphics[width=0.48\textwidth]{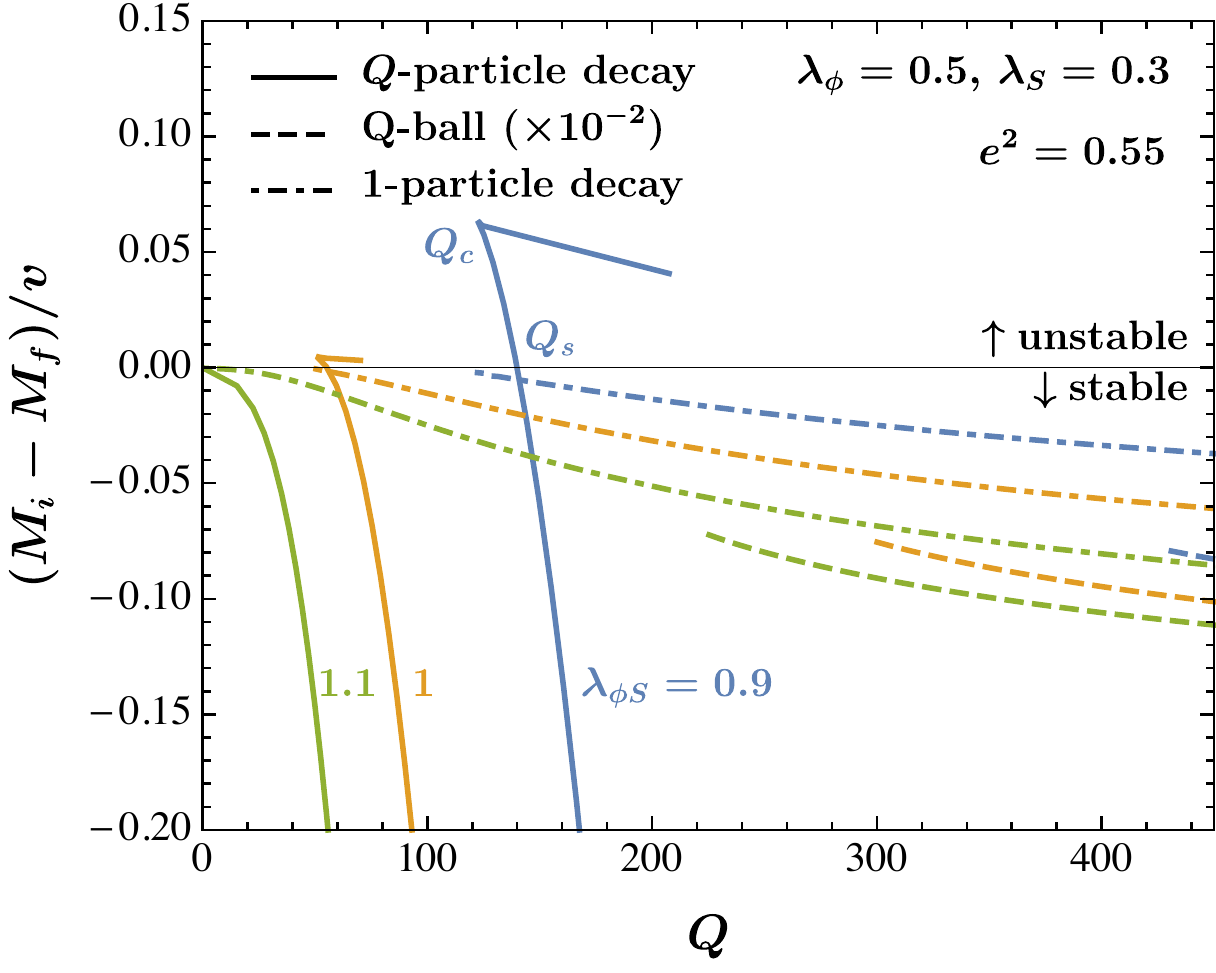}
	\caption{Difference between the initial and final masses $M_i$ and $M_f$ (normalized to $v$) for the $Q$-particle (\ref{eq:decay-to-free}), Q-ball (\ref{eq:decay-to-Qball}), and 1-particle (\ref{eq:decay-to-single-S}) decay channels shown solid, dashed, and dotted, respectively. Different colors denote different choices of $\lc$; the other parameters are fixed, including $\ms=0$.
		Values greater than zero are unstable; the charge at which the Q-particle decay crosses zero is denoted $Q_s$ in the text. When curves are cut off on the left, no solution exists for the QMB (solid and dotted curves) or for the isolated Q-ball (dashed curve), denoted $Q_c$ in the text. Two different-mass and same-charge solutions exist; part of the higher-energy solution for the QMB is shown in the solid curves. The Q-ball decay channel (dashed) has been multiplied by $10^{-2}$ on the y-axis to make it visible in the plot range. 
}
	\label{fig:mass-diff}
\end{figure}

With these solutions, we can numerically compute their masses using (\ref{eq:mass}), and compare to the masses of other states to numerically assess their stability. 
The difference of the initial and final state masses for each of the possible decay channels in (\ref{eq:decay-to-free}--\ref{eq:decay-to-single-S}) are compared in Fig.~\ref{fig:mass-diff} for a few choices of parameters as a function of $Q$. When all three channels have mass differences less than zero for a given charge, the QMB is stable. The largest $S$ charge $Q$ for which any of these channels' mass differences crosses zero is the minimum stable charge $Q_s$. In all cases in our full parameter scan, the decay to $Q$ free $S$ particles (\ref{eq:decay-to-free}) is the most important channel for determining $Q_s$. The green lines ($\lc=1.1$) in Fig.~\ref{fig:mass-diff} show an example with $Q_s=0$, so QMBs with any charge $Q$ are stable.

There is also a cutoff charge $Q_c$ below which no QMB or isolated Q-ball solutions (stable or unstable) exist. Recall that decreasing $Q$ corresponds to increasing $\Omega$. However, when $\Omega$ is increased sufficiently, the charge begins to increase again from $Q_c$, but the solutions are at a higher energy \cite{Friedberg:1976me}.  A portion of these higher-energy solutions for the QMBs are shown in the solid curves of Fig.~\ref{fig:mass-diff}, above the kinks. $\Omega$ is increasing from right to left along the lower branch, then continues from left to right along the upper branch as $\Omega$ approaches $\sqrt{\lc/2}$. The higher-energy solutions are unstable, so we will only focus on the lower-energy branch. A similar effect occurs for isolated Q-balls \cite{Ponton:2019hux}, but for the decays to isolated Q-balls we have cut off the corresponding dashed lines at $Q_c$ for the isolated Q-balls, omitting the higher-energy branch. In Q-ball literature, these higher-energy solutions are referred to as ``Q-clouds.'' Notice that $Q_c$ and $Q_s$ for the QMB are always smaller than $Q_c$ for the isolated Q-ball for fixed choice of Lagrangian parameters. This is because there is an additional energy savings from having the Q-ball and monopole overlap. The tunneling rate for quantum-mechanically unstable Q-balls with $Q_c < Q < Q_s$ in the lower energy branch to decay to free particles was considered in \cite{Levkov:2017paj}.

\begin{figure}[t]
	\centering
	\includegraphics[width=0.31\textwidth]{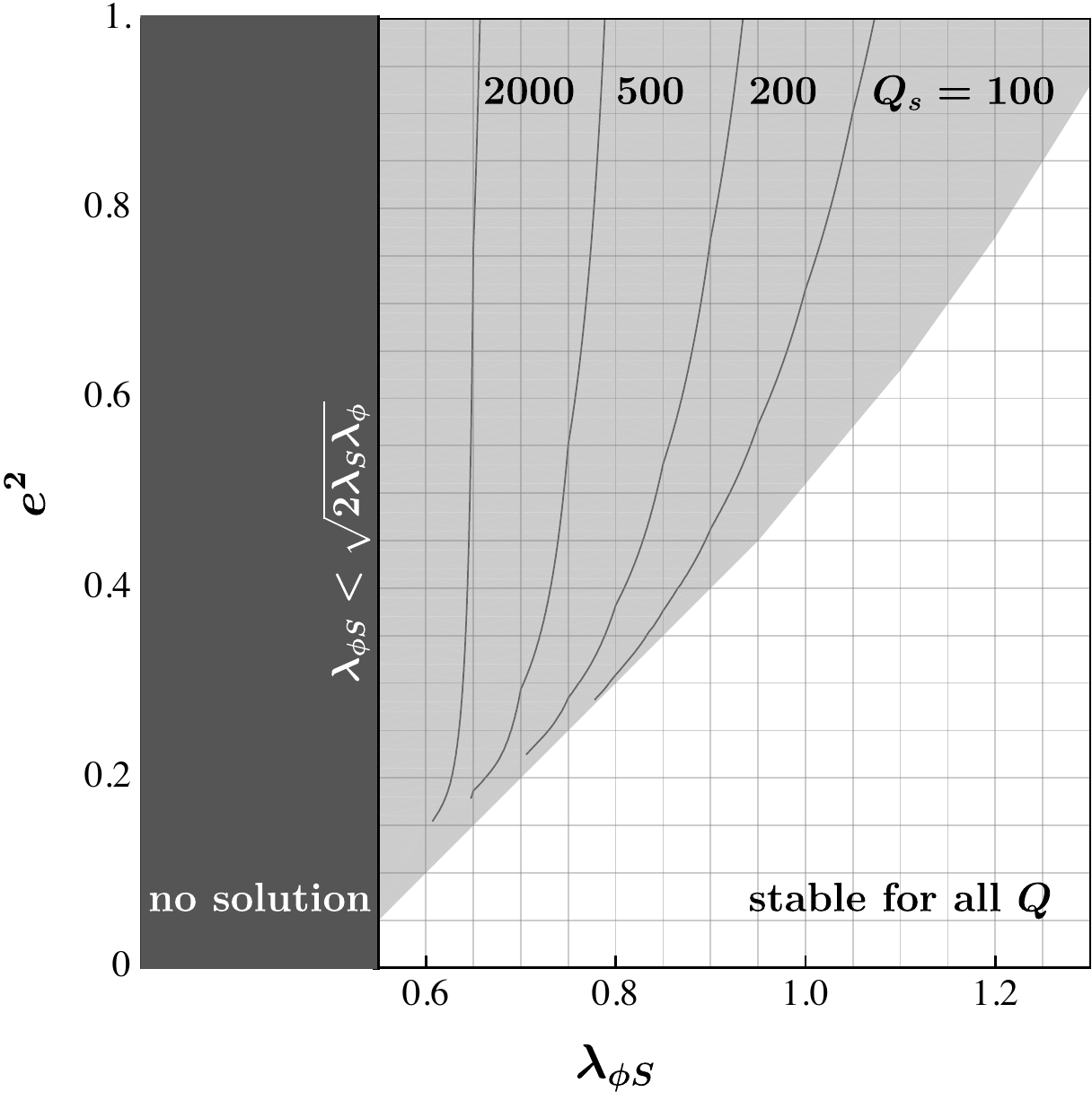} \hspace{3mm}
	\includegraphics[width=0.31\textwidth]{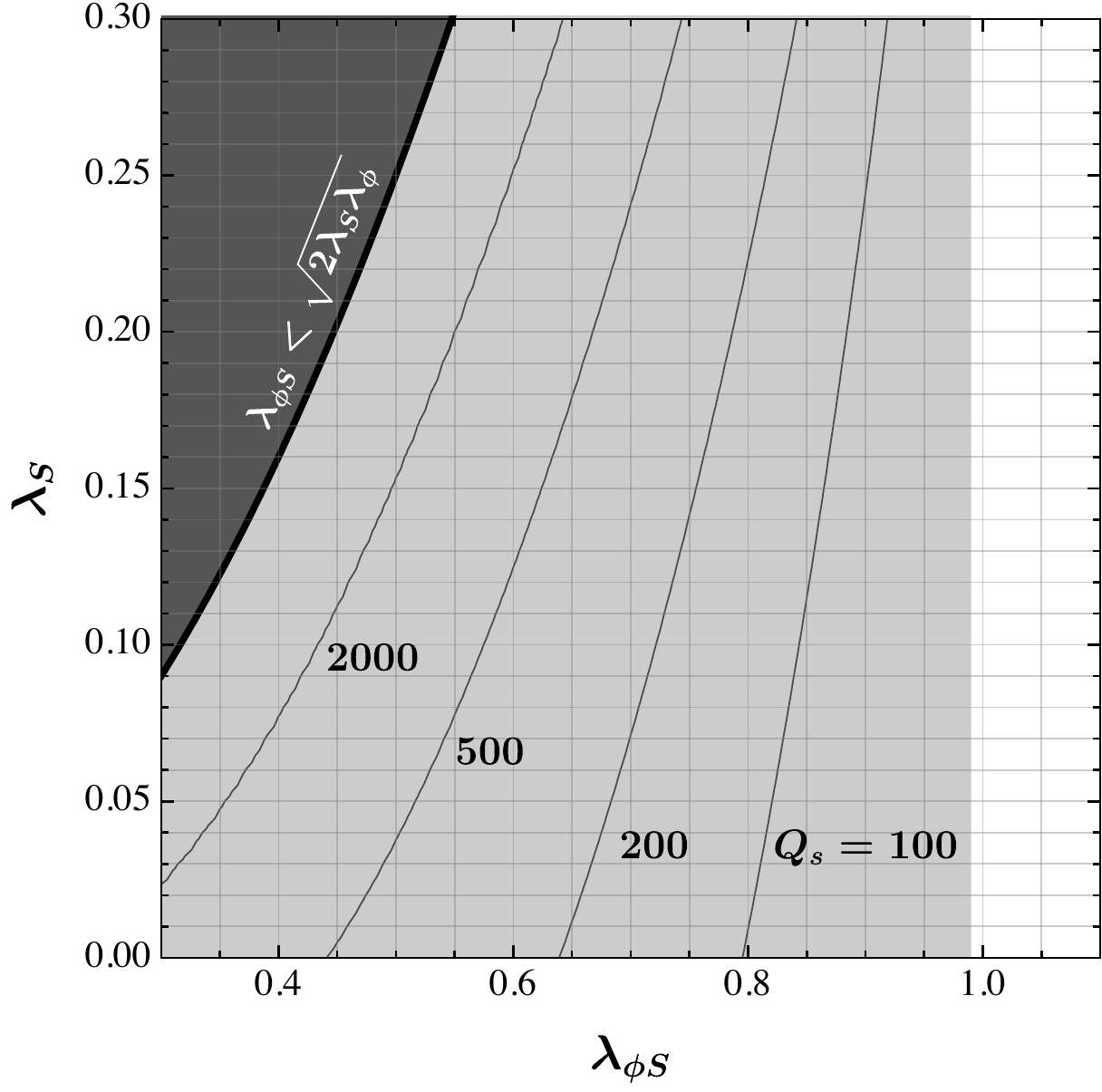} \hspace{3mm}
	\includegraphics[width=0.31\textwidth]{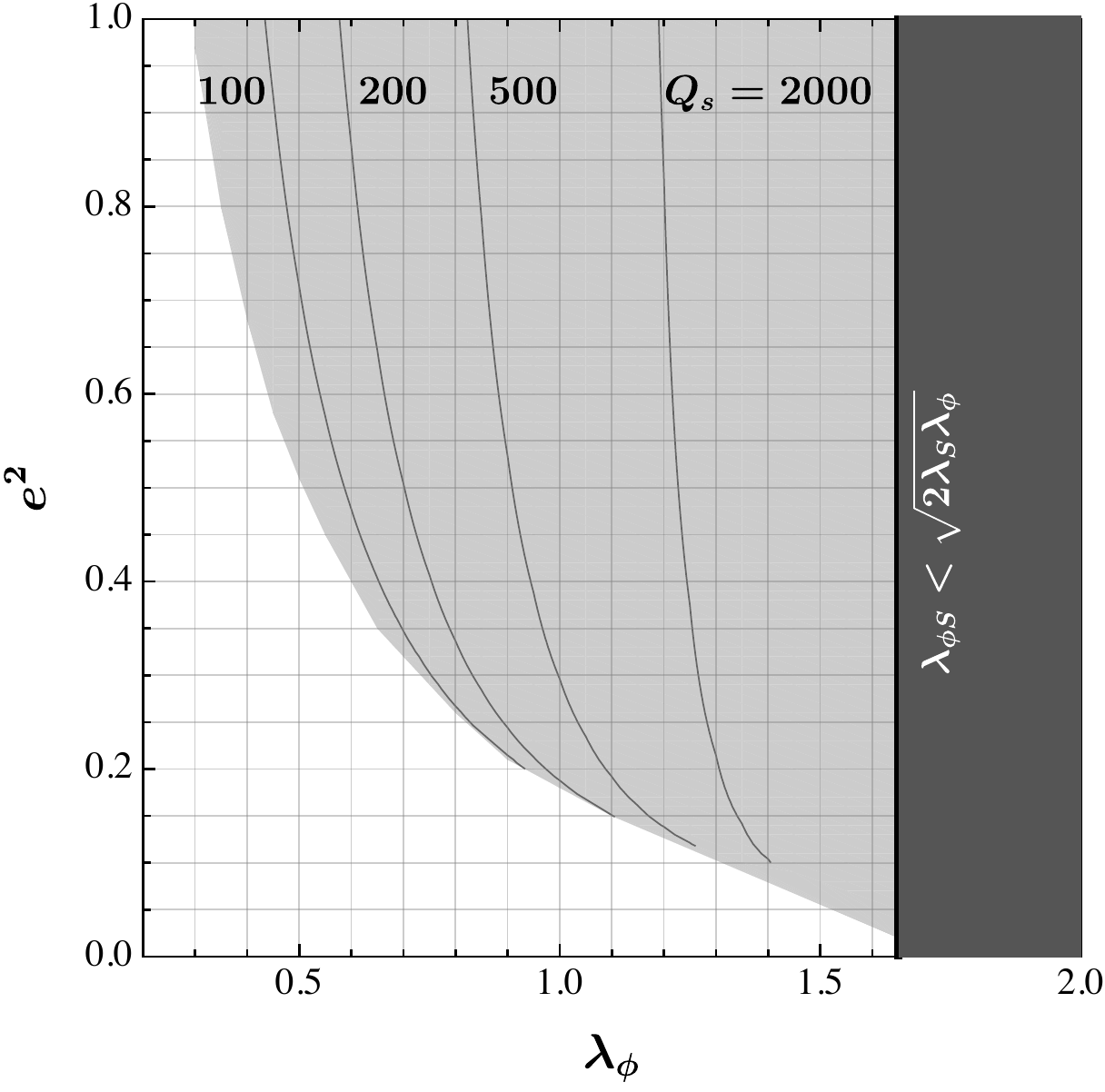}
	\caption{Contours of the minimum stable $S$ charge $Q_s$ for QMBs with magnetic charge $q=2$. When a Lagrangian parameter is not varied on the $x$- or $y$-axis, it is fixed to $\lphi=0.5$, $e^2=0.5$, $\ls=0.3$, $\lc=1$, and $\ms=0$. In the dark shaded regions, no Q-ball state exists of any kind, mixed QMB or isolated Q-ball. In the light shaded regions, $Q>Q_s$ is required for a stable solution. For $Q<Q_s$, the states decay to $Q$ free $S$ particles and an isolated monopole. The contours in the left and right panels should extend all the way to the dark shaded regions but are truncated due to limited numerical precision. In the white unshaded regions, all $S$ charges $Q$ down to unity are stable. }
	\label{fig:Qs-lc-e2}
\end{figure}

We show contours of $Q_s$ for example parameter choices in Figs.~\ref{fig:Qs-lc-e2}. In the white regions, all values for $Q$ down to $Q=0$ are stable (though in the quantized theory only integer $Q$ are allowed). As explained in Sec.~\ref{sec:small-Q}, our numerical scans confirm that when $\lphi=0$, all charges $Q$ provide stable mixed states, \ie, $Q_s=0$. However, for $\lphi>0$, there is a minimum stable $U(1)_S$ charge $Q_s$ below which mixed QMBs are unstable to decays to $Q$ free $S$ particles.  In the darkest shaded region no QMB or Q-ball solution is present. This region is only present when $\ls \lphi >0$ due to Eq.~(\ref{eq:lc-min}), otherwise $\lc$ can be arbitrarily small. In the intermediate lightly shaded region, QMBs are only stable if $Q>Q_s$. It is interesting to note that these stability threshold values are always smaller than the threshold $Q_s$ for an isolated Q-ball. For example, for the benchmark values in Fig.~\ref{fig:mass-diff}, $\lphi=0.5$, $e^2=0.55$, $\ls=0.3$, $\lc=1$, and $\ms=0$, $Q_s \approx 55$ for a QMB with $q=2$ while $Q_s \approx 345$ for a Q-ball with $q=0$.

Other patterns are also evident. Decreasing $e$ leads to increased stability for small $Q$ because the monopole radius is increased. The $S$ bound state energy is reduced because it is in a larger-radius potential well. Increasing $\lc$ also leads to increased stability for small $Q$ because a smaller $s_0$ is required to restore the non-Abelian gauge symmetry associated with $\phi$, which requires $\lc s_0^2 > \lphi v^2 / 2$. In the middle panel, the boundary between the white and light shaded regions is vertical because the effect of $\ls$ on the QMB mass is negligible in the small-$Q$ limit, as discussed in Sec.~\ref{sec:small-Q}.

Let us briefly discuss the case $\ms>0$ in relation to these numerical results, which all took $\ms = 0$. In Fig.~\ref{fig:Qs-lc-e2}, the boundaries between the dark shaded, light shaded, and white regions do not change when $\mus$ is varied. The boundary for the dark shaded regions is given by (\ref{eq:lc-min}), whose derivation held for arbitrary $\mus$. The boundary between the white and light shaded regions is determined in the $Q \to 0$ limit. In this limit, the mass contribution from nonzero $\mus$ in (\ref{eq:mass-mS0}) goes to zero proportional to $Q$. Thus, if a stable solution exists when $\mus=0$ for arbitrarily small $Q$, a stable $Q \to 0$ solution also exists for $\mus>0$ when holding all other Lagrangian parameters fixed. Therefore, only the contours of $Q_s>0$ shown in the light shaded regions of Fig.~\ref{fig:Qs-lc-e2} are modified by varying $\mus$. How these $Q_s$ contours vary with $\mus$ can be described by replacing $\lc$ with $\lbc$ in (\ref{eq:lbc}). Because $\lbc$ is a monotonically decreasing function of $\mus$, the value for $Q_s$ increases as $\mus$ increases; \ie, the QMBs become less stable as $\mus$ increases.

\section{Multiply magnetically charged QMBs ($q>2$)}
\label{sec:large-q}

We can extend the conclusions from the analysis of unit magnetically charged Q-balls to those with higher magnetic charges. It is well-known that spherical solutions for monopoles do not exist above unit monopole charge \cite{Weinberg:1976eq}. Thus, we cannot easily apply the numerical methods of Section \ref{sec:unitcharge} to this case. Nevertheless, we can make similar analytic approximations to argue for the stability of certain multiply charged configurations.

The case that most easily admits higher monopole charges is the case where the Q-ball radius is much larger than that of an isolated monopole. Then, as described in Sec.~\ref{sec:large-Q}, the QMB state can be thought of as a monopole bound inside a Q-ball. This modifies the energy in two ways. First, there is a vacuum energy savings because the interiors of both the monopole and Q-ball have $\phi=0$, which contributes to the vacuum energy in each system (when $\lphi > 0$). The volume of the QMB is smaller than the sum of the volumes of the isolated monopole and Q-ball, so the total vacuum energy is less for a QMB. We expect that adding additional monopoles to a QMB also has a negligible effect on the QMB radius, as long as $q$ is not too large compared to $Q$ that it significantly backreacts on the $s$ field in the equations of motion. 
Second, there is a change to the energy contained in the magnetic field. An isolated monopole's magnetic field energy is proportional to $v/e$. Once a monopole is bound inside a Q-ball, the magnetic charge spreads out throughout the QMB's volume (see the profile for $a$ in Fig.~\ref{fig:field-config} right panel). Thus, the energy contained in the field is proportional to $v/ (e^2 \rbc)$ as in (\ref{eq:M-step-contributions}). As a result, there is a reduction of the magnetic field energy when a single monopole (with characteristic radius $\rb \sim e^{-1}$ for $a(\rb)$) is trapped in a larger-radius Q-ball with radius $\rbc > e^{-1}$. This argument remains true as the monopole charge $q$ of the Q-ball increases, with the energy in the magnetic field going proportional to $q^2 v/ (e^2 \rbc)$. Only once $q$ is very large can the magnetic field energy of the $(q,Q)$ state be larger than the magnetic field energy for separated $(2,0)+(q-2,Q)$ states. Thus, $q$-charged QMBs should be stable up until some maximum value for $q$.

To be a bit more quantitative, we can generalize the approximation for the QMB mass in (\ref{eq:M-step-contributions}) to arbitrary magnetic charge $q$ as
\begin{equation}
\label{eq:M-step-contributions-q}
M_{(q,Q)} \sim \frac{304 \pi v (q/2)^2}{35 e^2 \rbc} + \frac{4 \pi}{3} \rbc^3 v \left(\frac{1}{4} \ls s_0^4 + \frac{1}{8} \lphi + \frac{1}{2} \Omega^2 s_0^2 \right) \, .
\end{equation}
While this does not capture the full nonspherical structure of the solution, it should give a reasonable approximation in the $Q,\rbc \gg 1$ limit. In this limit, the first term is negligible compared to the second term as we noted previously, and the other terms set $\rbc$, $s_0$, and $M$ as a function of $Q$ as in (\ref{eq:uc-largeQ}--\ref{eq:g0-largeQ}).

Using the notation from Eqs.~(\ref{eq:decay-to-free}--\ref{eq:decay-to-single-S}), in order for the bound $(q,Q)$ state to be stable against decay to $(q, Q) \rightarrow (q-2,Q)+(2,0)$, there is an upper bound on $q$ as~\footnote{The condition that the first term in (\ref{eq:M-step-contributions-q}) is negligible is $q \ll \ls^{1/6} e \, Q^{2/3}$, less important than the condition in (\ref{eq:qmax}) for large $Q$.}
\beqa
q \lesssim 1+ \frac{35}{76}\, e\, \rbc\, Y   \approx \frac{35 (3/\pi)^{1/3}}{76 \times 2^{5/12}} e \, Y \frac{\ls^{1/12}}{\lphi^{1/4}} Q^{1/3} \, ,
\label{eq:qmax}
\eeqa
where in the last step we have substituted (\ref{eq:uc-largeQ}), which is only valid when both $\lphi,\ls>0$. The other possible decay channel to $(q/2) \times (2,0) + (0,Q)$ is less important, having a larger upper bound on $q$ by a factor of 2 than the one above. The binding energy to add an isolated $(2,0)$ monopole to an existing $(q,Q)$ state to form a $(q+2,Q)$ state is the same as (\ref{eq:binding-single-monopole}) but with the last term on the right-hand side multiplied by $(q+1)$. In the limit of $(q +1) \ll e\,\ls^{1/12}\lphi^{-1/4}Q^{1/3}$, the binding energy saturates the isolated $q=2$ monopole mass $\Delta M \approx M_{(2,0)}$.

This approximation supports our claim that mixed $(q,Q)$ solutions exist and are stable for a sufficiently large hierarchy between $q$ and $Q$. While more detailed numerical work is largely intractable, the conclusion remains that large-charged monopoles may exist in nature.

\begin{figure}[t!]
	\centering
	\includegraphics[width=0.6\textwidth]{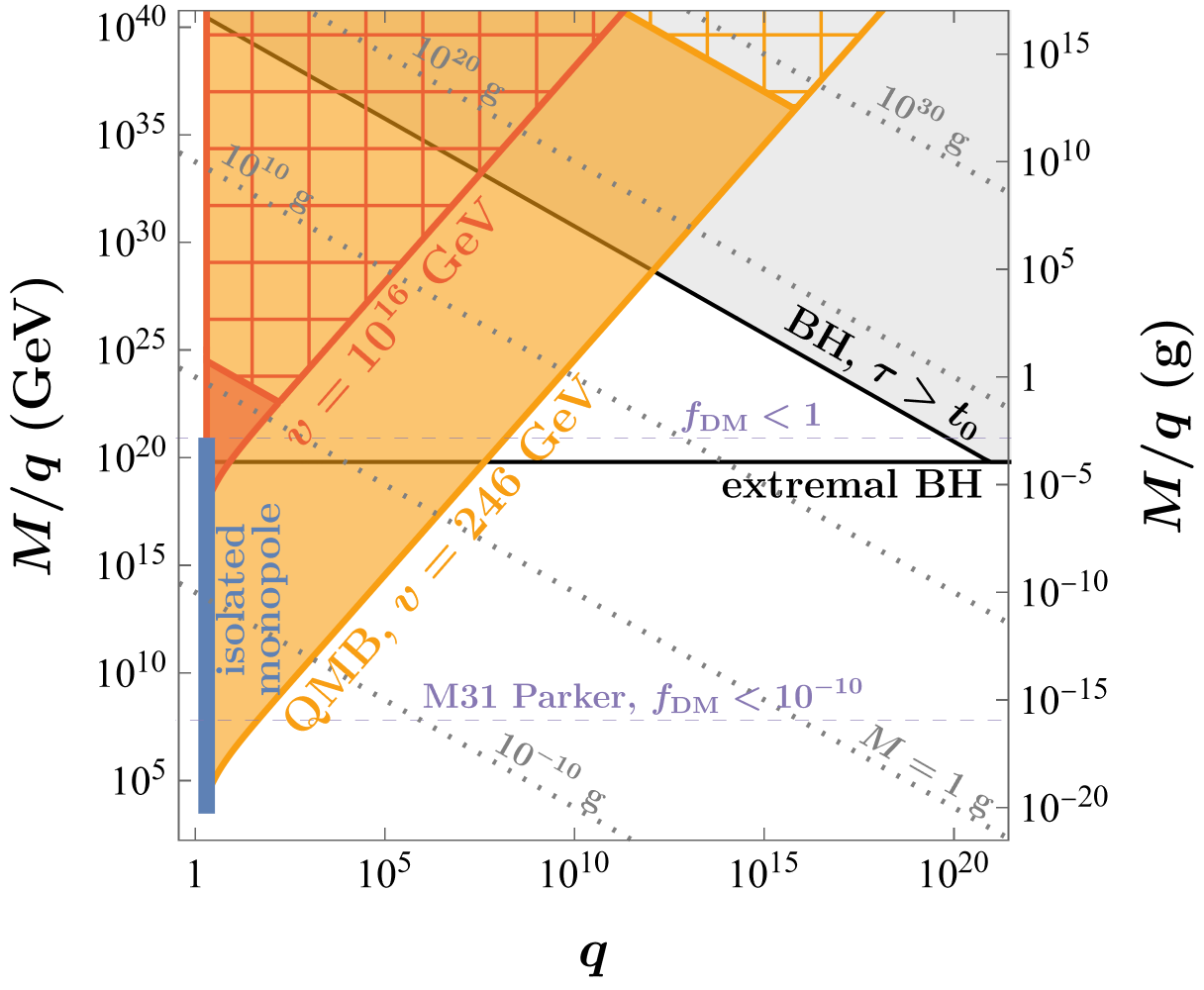}
	\caption{Allowed mass-to-charge ratio $M/q$ and magnetic charge $q$ for various cosmologically stable magnetically charged states. Orange and red regions: QMBs with $\lphi=0.5$, $\ls>0$ (insensitive to exact value), $e^2 = 4 \pi / 137$ the SM value, and $v=246~(10^{16}) \, \GeV$ for orange (red). The bottom-left boundaries of each region with small $q$ and $Q$ are not reliable because the large $Q$ limit is assumed. Hashed regions show where the Schwarzschild radius exceeds the QMB radius, so gravitational effects (ignored here) should be taken into account. Blue vertical line: isolated monopoles with $e$ the SM value and $v$ varying from 246 GeV to $\Mpl$, using the same $\lphi$ and $e$ as the QMB. Black: magnetically charged black holes, which exist as cosmologically stable relics today either on the horizontal black line if they are extremal or in the shaded region if they are not extremal (in between, they evaporate to the extremal line with lifetime $\tau \ll t_0$ the age of the universe). Lilac horizontal dashed lines indicate the bound on the relic abundance as a fraction of the dark matter density $f_\text{DM}$ from the Parker bound applied to M31 (assuming all relics have the same charge and mass). Above the line labeled ``$f_\text{DM}<1$,'' magnetically charged states could explain all of dark matter (though other model-dependent bounds exist). Gray dotted lines show contours of constant mass.}
	\label{fig:MvsQmag}
\end{figure}

QMBs thus offer a wider variety of charges and masses for magnetically-charged objects than other systems. For an ordinary isolated monopole, the mass and charge are fixed to $M = 4 \pi v Y / e$ and $q=2$, respectively.
On the other hand, a QMB will have mass given approximately by (\ref{eq:M-largeQ})  and magnetic charge satisfying (\ref{eq:qmax}). Thus,
\begin{equation}
M_{(q,Q)}
\gtrsim 21 \frac{\lphi}{e^3 Y^3} q^3 v \, .
\label{eq:mass-large-q}
\end{equation}
Again, this is only valid when $Q \gg 1$ and both $\lphi, \ls>0$. This could also be contrasted with magnetically charged black holes (BH) \cite{Lee:1994sk,Maldacena:2020skw,Bai:2020spd,Bai:2020ezy}, another example of a bound state admitting larger $q>2$. If lighter than about $10^{17} \, \text{g}$, they evaporate to extremal very quickly with lifetime $\tau \ll t_0$ the age of the universe, at which point they become cosmologically stable \cite{Bai:2019zcd}.~\footnote{If the magnetic charge $q$ is sufficiently small---much less than extremal, so that they do not have an enhanced Hawking evaporation rate---magnetic BHs down to about $10^{15} \, \text{g}$ are cosmologically stable. However, evaporation products of PBHs below $10^{17} \, \text{g}$ set limits on their abundance (for review, see \cite{Carr:2020gox}). Thus, while the ``BH, $\tau>t_0$'' line in Fig.~\ref{fig:MvsQmag} would curve slightly downwards as it approaches smaller $q$, we neglect this because such BHs could only be a small fraction of DM.}
Extremal black holes have a fixed mass-to-charge ratio $M/q =\cos\theta_W \sqrt{\pi} e^{-1} \Mpl \approx 5.2 \, \Mpl$ with $\Mpl$ the Planck mass and $\theta_W$ the weak mixing angle. Their masses also cannot be smaller than the Planck scale. 

The possible masses and magnetic charges for various magnetically charged states are compared in Fig.~\ref{fig:MvsQmag}. Isolated monopoles (blue line) can only have charge $q=2$ and their mass depends on $v$. QMBs can have larger charge and mass than isolated monopoles; their mass is always larger than that of isolated monopoles. The boundary from (\ref{eq:mass-large-q}) is derived in the large $Q$ limit and thus should not be trusted in the small $q$ and $Q$ limit (bottom left of both the orange and red regions). Magnetic black holes may either be near-extremal (along the horizontal black line) or sufficiently massive to have not evaporated to near-extremal (shaded region). In comparison, QMBs allow a larger variety of masses for a given magnetic charge. QMBs can have large magnetic charges with masses below the Planck scale, not allowed for black holes. QMBs can also exist in the parameter region where magnetic BHs would have evaporated to extremal (see Fig.~\ref{fig:MvsQmag}, between the black horizontal line and black shaded region).

All of our derivations so far have neglected gravitational effects. However, once the Schwarzschild radius $r_s = 2 M/\Mpl^2$ is larger than the QMB radius, we expect such effects to be important, potentially leading to a soliton star or black hole \cite{Lee:1986ts,Friedberg:1986tq,Lee:1991ax}. Comparing $r_s$ to (\ref{eq:uc-largeQ}), gravitational effects are important when
\begin{equation}
M_{(q,Q)} \gtrsim \sqrt{\frac{3}{16\pi}} \, \lphi^{-1/2} \left(\frac{\Mpl}{v}\right)^2 \Mpl \, .
\end{equation}
This is denoted by the hatched region in Fig.~\ref{fig:MvsQmag}. Larger $v$ leads to more compact QMBs, so gravitational effects are more important.

\section{Early-universe formation of QMBs}
\label{sec:formation}
       
The QMB could be formed during a phase transition in the early universe at a temperature $T \sim v$. For example, if the phase transition is first order, the true-vacuum bubbles could ``snowplow" the global charge into a small pocket and form the Q-balls~\cite{Ponton:2019hux,Witten:1984rs}. The average global charge for the Q-balls depends on the symmetry breaking scale $v$, the rate of bubble nucleation for the phase transition, and the potentially nonzero initial matter-antimatter asymmetry for the $S$ particles $\eta_Q=|n_S - n_{S^\dagger}|/(n_S + n_{S^\dagger})$ where $n_{S^{(\dagger)}}$ is the $S$ (anti)particle number density. The QMB's acquisition of magnetic charge is trickier and relies on the topological structure of the field configuration. The Kibble mechanism is usually adopted to estimate the formation abundance for a unit magnetic charge with $q=2$. The probability to form a large winding number is usually suppressed, which will be also discussed later. 

Let us consider a first order phase transition based on the finite-temperature potential of  $\phi^a$. We begin with the nontopological global charge, then later discuss the topological magnetic charge. The global charge of QMBs and Q-balls relies on the total number of $S$ particles within one Hubble patch during the phase transition and the number of nucleation sites at the end of the phase transition. The total number of $S$ particles (including both $S$ and $S^\dagger$) within one Hubble volume at the temperature $T_f$ when the phase transition ends is 
\beqa
N_S^{\rm Hubble} \sim n_S \,d_H^3 \sim (1 \times 10^{46})\, v_3^{-3} ~,
\eeqa
where $v_3 \equiv v/(10^3 \, \GeV)$, $1/d_H = H(T_f) = \sqrt{\pi^2 g_*/90}\,T_f^2/M_{\rm pl}$, and $n_S = (2 \zeta(3)/\pi^2) T_f^3$ the (relativistic) $S$ number density with $g_*  \approx 100$ and $T_f\sim v$. The number of bubble nucleation sites per Hubble patch is~\cite{Ponton:2019hux}
\beqa
N^{\rm Hubble}_{\rm QMB} \sim (1\times 10^{13}) \times \left( \frac{\lc}{3} \right)^{-14} ~,
\eeqa
where the large power $14$ comes from a numerical fit.
Note that for large enough $\lc \gtrsim 2$, the phase transition is strongly first order \cite{Katz:2014bha,Jain:2017sqm,Ponton:2019hux}. So, the average number of both $S$ and $S^\dagger$ particles in one QMB is
\beqa
N^{\rm QMB}_{S} \sim (1 \times 10^{33}) \, v_3^{-3}\, \left( \frac{\lc}{3} \right)^{14} ~,
\eeqa
assuming most or all of the $S$ particles remain in the false vacuum.
Taking the particle-antiparticle annihilations into account, the typical $Q$ charge for each QMB is (taking $\eta_Q \ll 2$)
\beqa
\label{eq:QMB-charge}
\langle Q_{\rm QMB} \rangle \sim \mbox{max}\left[(3 \times 10^{16})\, v_3^{-3/2}  \left( \frac{\lambda_{\phi S}}{3} \right)^{7} , \quad \eta_Q\,(1 \times 10^{33})\, v_3^{-3}  \left( \frac{\lambda_{\phi S}}{3} \right)^{14}  \right] ~.
\eeqa
The second term comes from if the asymmetric component dominates. Otherwise, the first term gives the typical net charge from the symmetric component $Q \sim (N^\text{QMB}_{S})^{1/2}$. Since the net charge in this equation is typically large, the QMB's mass is dominated by the Q-ball part, and the average mass of QMBs at their formation is $\langle M_{\rm QMB} \rangle \approx \langle Q_{\rm QMB} \rangle \Omega_c v$.

If QMB annihilations are not important, the yield of QMBs $Y_\text{QMB} \sim N^\text{Hubble}_\text{QMB} d_H^{-3} s^{-1}$, with $s = (2\pi^2/45)g_{*S} T_f^3$ and $g_{*S} \approx 100$, is constant. It can be compared to the observed dark matter abundance $\Omega_\text{DM} h^2 \approx 0.12$ \cite{Planck:2018vyg}, which corresponds to $Y_\text{DM} \approx 3.6 \times 10^{-10} (1 \, \GeV / M_\text{DM})$. Thus, QMBs and Q-balls would make up a fraction $f_\text{DM}$ of dark matter
\begin{equation}
\label{eq:fDM-firstorder}
f_\text{DM} \sim \max \left[ (4 \times 10^{-7}) v_3^{5/2} \left( \frac{3}{\lambda_{\phi S}} \right)^{7} , \quad \eta_Q (2 \times 10^{10}) v_3 \right] \Omega_c \, .
\end{equation}
For QMBs and Q-balls to account for all the dark matter abundance
\beqa
\left\{ \begin{array}{l } \mbox{asymmetric component dominates:} \\
\qquad \eta_Q \sim (6 \times 10^{-11} v_3^{-1}) \Omega_c^{-1}   \quad \mbox{and} \quad v \lesssim (3\times 10^5 ~\mbox{GeV}) \left(\frac{\lambda_{\phi S}}{3} \right)^{14/5} \Omega_c^{-2/5}
\vspace{3mm}
\\ 
\mbox{symmetric component dominates:} \\
\qquad v \sim (3\times 10^5 ~\mbox{GeV}) \left(\frac{\lambda_{\phi S}}{3} \right)^{14/5} \Omega_c^{-2/5} \quad \mbox{and} \quad \eta_Q \lesssim (2 \times 10^{-13}) \Omega_c^{-1} \left( \frac{3}{\lambda_{\phi S}} \right)^{14/5} \end{array} \right. \, .
\eeqa

The upper line corresponds to the asymmetric component dominating $\langle Q_\text{QMB} \rangle$, while the lower corresponds to statistical fluctuations in the symmetric component dominating (the second and first terms, respectively, of the $\max$ functions in (\ref{eq:QMB-charge}--\ref{eq:fDM-firstorder})). Note that $\eta_Q$ may be similar to the baryon asymmetry $\eta_B \approx 6 \times 10^{-11}$, suggesting a potential shared asymmetry formation mechanism for both baryons and $S$ (to which we remain agnostic here). Thus, the typical QMB and Q-ball mass if they explain all of dark matter ($f_\text{DM}=1$) is
\begin{equation}
\langle M_{\rm QMB} \rangle \sim  (8 \times 10^{25}\,\mbox{GeV})\,v_3^{-3} \, \left( \frac{\lambda_{\phi S}}{3} \right)^{14} \quad \mbox{for} \quad v\lesssim (3\times 10^5 ~\mbox{GeV}) \left(\frac{\lambda_{\phi S}}{3} \right)^{14/5} \Omega_c^{-2/5}  ~.
\end{equation}
The upper limit on $v$ corresponds to a lower limit $\langle M_{\rm QMB}\rangle \gtrsim (2\times 10^{18}\,\mbox{GeV})  (\lambda_{\phi S}/3)^{28/5} \Omega_c^{6/5}$.

Topological defects form where multiple bubbles with different symmetry-breaking phases meet. In a first order phase transition, these topological defect formation sites coincide with where Q-balls form due to the snowplow mechanism---both form where multiple bubbles meet. Thus, where a monopole is formed, it is almost guaranteed to be already inside a Q-ball. This is true as long as the typical charge $\langle Q_{\rm QMB} \rangle \gg Q_s$, which should generally be the case for the large charges in (\ref{eq:QMB-charge}). Conversely, not every site where a Q-ball is formed will have a monopole. Most QMBs formed in such a manner will have initial magnetic charge $q=2$.  The probability for a larger winding number $q>2$ for QMBs is suppressed for geometric reasons. The number of true vacuum bubbles surrounding the pocket of false vacuum which forms a QMB determines the probability of the topological winding number. As a simple example, in two spatial dimensions, a vortex is formed with probability 1/4 when three bubbles meet. Three bubbles cannot form a vortex with larger than unit winding number---at least five bubbles must meet to form winding number two, and so forth for higher windings. Even if five bubbles meet, the probability to form winding number two is only $\sim 0.005$. Similarly, for three spatial dimensions, the probability for four bubbles (in a tetrahedron shape) to form a unit winding number is 1/8, and higher monopole charges require the confluence of more bubbles~\cite{Prokopec:1991ab,Leese:1990cj,Leese:1991gt,Okabe:1999is}. The probability to form higher monopole charges $q>2$ is thus suppressed by both the probability for many ($>4$) bubbles to collide at approximately the same point and the probability for those bubbles to produce a higher winding number. For simplicity, we parameterize the probability of Q-balls to carry a unit magnetic charge to be $p_2$ between about $10^{-1}$ and $10^{-2}$. So, at the end of the phase transition, there are a fractional abundance of $1-p_2$ Q-ball states and $p_2$ QMB states with magnetic charge $q=2$. The fraction of QMBs carrying a larger $q$ and the fraction of isolated monopoles without a global charge are suppressed.
       
After the formation of Q-balls and QMBs, the charge distributions could continue to evolve via the so-called ``solitosynthesis"~\cite{Griest:1989bq}. This can proceed in two ways: (a) Q-balls and QMBs can merge if they have same-sign global charge or (partially) annihilate if opposite-sign, and (b) Q-balls and QMBs can absorb unbound $S$ particles. For a sufficiently strong first order phase transition corresponding to larger $\lc$, (b) can be approximately neglected because almost all $S$ particles will be bound inside Q-balls or QMBs \cite{Ponton:2019hux}. Further, a simple estimate reveals (a) to be unimportant. Assuming for simplicity that all Q-balls and QMBs have initial nontopological charge $Q$, the rate for two such objects to merge or annihilate is $\Gamma = \sigma\, v_{\rm rel} n_Q$ with $\sigma \sim \pi R_Q^2$ and $R_Q \sim Q^{1/3}\,v^{-1}$. The relative speed is $v_{\rm rel} \sim \sqrt{T/M_Q}$, and the number density is $n_Q \sim T_{\rm eq} T^3/M_Q$ with $T_{\rm eq} \approx 1$~eV as the temperature of matter-radiation equality and assuming Q-balls and QMBs account for all dark matter or $f_{\rm DM} = 1$. Comparing $\Gamma$ to the Hubble rate $H(T) \sim T^2/M_{\rm pl}$, the interaction rate is important when $Q < 6\times 10^4\, v_3^{-12/5}$ with $T \sim v$. Comparing this to the typical charge in \eqref{eq:QMB-charge}, the additional charge evolution effects (in both $q$ and $Q$) are not important in this formation mechanism. Here, we have taken the geometric cross section of the QMBs for $\sigma$. Note that the cross section for magnetic monopole interactions is generally much larger and plays a role in annihilations of isolated monopoles \cite{Preskill:1979zi}. However, we have verified that this effect is unimportant for QMBs carrying small $q \sim 2$ and large mass $M_{(q,Q)} \gg M_{(2,0)}$.

If the phase transition is second order, Q-balls can form in finite regions of false vacuum. The probability for a region to remain in the false vacuum is determined by the Boltzmann distribution comparing the region's free energy to the Ginzburg temperature (the temperature at which thermal fluctuations between the false and true vacua freeze out) \cite{Frieman:1988ut,Griest:1989cb,Frieman:1989bx}. Then, Q-balls form in these false vacuum regions either due to an initial asymmetry $\eta_Q$ or statistical fluctuations in the difference of $S$ and $S^\dagger$ particles, similar to the first order phase transition. Meanwhile, monopoles can form isolated from Q-balls via the Kibble-Zurek mechanism \cite{Zurek:1985qw,Murayama:2009nj}. Unlike the first order phase transition, the $S$ particles are not snowplowed to the boundaries between different coherent regions of symmetry-breaking phases, so Q-balls and monopoles need not initially overlap. Additionally, it cannot be assumed that few free $S$ particles remain unbound after the phase transition. Thus, following the phase transition, the charges of the states may evolve. If $Q_s$ is sufficiently small, the monopoles could form bound states with free $S$ particles, building up their $U(1)_S$ charge $Q$ either due to an asymmetry in $S$ particles or due to fluctuations in a 1-dimensional random walk in $Q$. Importantly, because $Q_s$ can be arbitrarily small for QMBs (contrary to Q-balls), it is easier to build QMBs from fusion of free $S$ particles than it is to build Q-balls, the latter of which requires many $S$ particles ($>Q_s$) to meet simultaneously \cite{Griest:1989bq,Frieman:1989bx}. Thus, solitosynthesis can more easily produce a population of QMBs. Additionally, such monopoles or QMBs could encounter isolated Q-balls and become bound by them. Q-balls and QMBs could also be destroyed by free (anti)particles if their charge is annihilated below $Q_s$ \cite{Griest:1989bq}. The details of the many types of interactions between free particles and bound states require tracking of interactions of the many different possible states, which is beyond the scope of this work.

\section{Implications for phenomenology}
\label{sec:pheno}

The Parker bound \cite{Parker:1970xv,Turner:1982ag} states that coherent magnetic fields cannot be drained of energy by accelerated monopoles, otherwise we would not observe them. When a monopole passes through a coherent magnetic field, it is accelerated to a speed
\begin{equation}
v_\text{mag} \simeq \min \left[ 1,\; \sqrt{\frac{2 B (2\pi q/e) \ell_c}{M}} \right] \simeq \min \left[ 1 ,\; 4 \times 10^{-5} \sqrt{\frac{q}{M} 5.2 \Mpl \ell_{21} B_3} \right] \, ,
\end{equation}
where $\ell_{21} = \ell_c/(10^{21}\,\text{cm})$ is the coherent length of the magnetic field with strength $B_3 = B/(3 \, \mu\text{G})$.
The Parker bound was initially derived for Milky Way coherent fields. It was recently extended to M31 \cite{Bai:2020spd}, which has larger-scale approximately axisymmetric magnetic fields in its disc with $\ell_{21} \sim 30$ \cite{Fletcher:2003ec} and a corresponding regeneration time $t_\text{reg}\sim 10 \, \text{Gyr}$ \cite{Arshakian:2008cx}. Its strength is $B_3 \sim 5/3$ \cite{Fletcher:2003ec}, and its local dark matter (DM) density and escape velocity are similar to the Milky Way: $\rho_{0.4} = \rho / (0.4 \, \text{GeV cm}^{-3})$ \cite{Klypin:2001xu,Tamm:2012hw} and $v \sim 10^{-3}$. The M31 Parker bound on the DM energy density fraction of QMBs $f_\text{DM} = \rho_\text{QMB}/\rho_\text{DM}$ thus takes the form
\beqa
f_\text{DM} \lesssim \left\{ 
\begin{array}{l l} 6 \times 10^{-3} \left(\dfrac{M}{5.2 \, q \, \Mpl}\right)^2 \dfrac{v_{-3}}{\rho_{0.4} (\ell_{21}/30) (t_{15}/300)} \, , & v_\text{mag} \lesssim 10^{-3} \vspace{4mm}
\\
10^2 \left(\dfrac{M}{5.2 \, q \, \Mpl}\right) \dfrac{B_3/(5/3)}{v_{-3} (\rho_{0.4}/10^{-6})(t_{15}/300)} \, , & v_\text{mag} \gtrsim 10^{-3}
\end{array} \right. \, .
\label{eq:parker}
\eeqa
Here, $v_{-3} = v/10^{-3}$ is the velocity of the QMBs. If $v_\text{mag} \lesssim 10^{-3}$, the QMBs can be bound in the galaxy with $v_{-3} \sim 1$ and could explain DM. Otherwise, they cannot be bound and must have $v_{-3} \gtrsim 1$; in that case, they cannot explain all of DM. If they are bound, $f_\text{DM}$ is compared to the local DM density $\rho_{0.4} \sim 1$; otherwise, if they are unbound $f_\text{DM}$ is compared to the average DM density $\rho_{0.4} \sim 10^{-6}$. These differences in $\rho$ are reflected in the top and bottom lines of (\ref{eq:parker}). The parameter $t_{15} = t_\text{reg} / (10^{15} \, \text{s})$. This bound is represented by lilac dashed lines in Fig.~\ref{fig:MvsQmag}.

The most constraining direct search for $q \geq 2$ monopoles comes from searches for tracks in ancient mica \cite{Ghosh:1990ki}, which limits $f_\text{DM} < (M / 10^{23} \, \GeV)$ (about an order of magnitude stronger than the strongest ``laboratory'' experiment MACRO \cite{MACRO:2002jdv}). Note, we expect the sensitivity of these searches is roughly insensitive to the magnetic charge $q$. They are slightly more constraining than the Parker bound for small $q$ but subdominant at large $q$.

Direct detection constraints could also apply independent of the magnetic charge, especially if the symmetry breaking in the full theory makes the QMB interior electroweak symmetric \cite{Ponton:2019hux,Bai:2019ogh,Bai:2020ttp,Adhikari:2021fum}. These are most stringent for smaller $v$ (near the electroweak scale) because the QMB radius goes as $1/v$ in the large-$Q$ limit. They can be completely alleviated at larger $v$, reducing the effective cross section.

Other model-dependent bounds on magnetically charged objects can be found in Refs.~\cite{Bai:2020spd,Mavromatos:2020gwk,Ghosh:2020tdu,Diamond:2021scl,Bai:2021ewf}, though things like stopping power in materials may need to be recalculated to apply some of these bounds to QMBs (especially if the QMB radius is large). Additionally, magnetic monopoles could be detected via their effects as they pass through the atmosphere. Previous atmospheric studies have focused on states with dipole rather than monopole magnetic fields \cite{VanDevender:2017avk,Sloan:2021pmf} or with a fixed geometric scattering cross section \cite{Sidhu:2019fgg} and would need to be reinterpreted accordingly.

We briefly mention a novel proposal for a detection strategy, although it is unlikely to compete against other established bounds unless undertaken at great scale. Magnetic anomaly detectors (MADs) \cite{Zhao_2020} offer another interesting method to search for passing magnetically charged particles like QMBs that has not been previously considered in literature to our knowledge. The advantage of such detectors is that they are commercially available, compact, and work at room temperature. They have typical sensitivities of order $S_\text{MAD} \sim 1~\text{to}~10^{3}~\text{fT}/\sqrt{\text{Hz}}$ (see, \eg, \cite{magnetometer,doi:10.1063/1.3491215,doi:10.1063/1.4905449,Li_2017}). A passing QMB, which generates a monopole magnetic field $B_m = (10^{-6} \, \text{fT}) (q/2) (\text{km}/r)^2$ at a distance $r$ from its center, could generate a signal in a MAD. If bound in the galaxy, its velocity is $v \sim 10^{-3}$, so the passing time for it to be within a distance $r$ from a MAD is $t_\text{pass} \sim (3 \times 10^{-3} \, \text{s}) (r/\text{km}) (v_{-3})^{-1}$. To be above the MAD threshold, $B_m \sqrt{t_\text{pass}} > S_\text{MAD}$ is required. Using the local density of dark matter, the number of QMBs that pass within a distance $r$ of a MAD during an exposure time $T$ is $N \sim f_\text{DM} (T/\text{s}) (10^{17} \, \GeV/M) (r/\text{km})^2 v_{-3}\, \rho_{0.4}$. Allowing there to be $N_\text{MAD}$ different MAD detectors in an experiment (spaced sufficiently far apart that their effective sensitive areas do not overlap), the number of detected QMBs $N_\text{det}$ above the MAD threshold sensitivity is thus
\begin{equation}
N_\text{det} \sim 7 \times 10^{-7} \, N_\text{MAD} \, f_\text{DM} \, q^{4/3} \, \left(\frac{10^{17} \, \GeV}{M} \right) \left(\frac{\text{fT}/\sqrt{\text{Hz}}}{S_\text{MAD}}\right)^{4/3} \left(\frac{T}{\text{yr}}\right) (v_{-3})^{1/3} \, .
\label{eq:Ndet}
\end{equation}
Additionally, MADs may lose sensitivity at small frequencies, for example of order Hz for SQUID detectors~\cite{ripka2021magnetic}. If we conservatively require $t_\text{pass} < t_\text{pass,max} = 1 \, \text{s}$, this limits the effective distance of closest approach to $r < 300 \, \text{km} \,(t_\text{pass,max}/\text{s})$. Then, there is a bound on $N_\text{det}$ going as
\begin{equation}
N_\text{det} \lesssim 4 \times 10^{12} \, N_\text{MAD} \, f_\text{DM} \left(\frac{10^{17} \, \GeV}{M} \right) \left(\frac{T}{\text{yr}}\right) \left(\frac{t_\text{pass,max}}{\text{s}}\right)^2 \, .
\end{equation}
Unlike (\ref{eq:Ndet}), this is independent of $q$, so it sets an upper mass reach. A search could be sensitive when $N_\text{det} \gtrsim 3$, assuming no backgrounds. To reduce backgrounds, a lattice configuration of MADs could be used to look for correlated signals of a passing QMB. A less sensitive network of over 100 detectors, accurate to the nT level, is already in place for monitoring the geomagnetic field \cite{doi:10.1063/1.2883907}.  Compared to other direct searches, this search strategy has the advantage that one detector can cover a large cross sectional area, especially for larger $q$. However, barring a very large number $N_\text{MAD}$ or very sensitive detectors with much smaller $S_\text{MAD}$, the Parker bound above tends to give a stronger constraint on $f_\text{DM}$. Note, these sensitivities do not account for the possibility that large-$q$ and small-$M$ QMBs could be stopped in the atmosphere.

\section{Conclusions}
\label{sec:conclusion}

In this work, we have established the existence of a new physical state carrying both topological and nontopological charge. In the analytically tractable case of unit magnetic charge $q=2$, we have solved the spherically symmetric equations of motion for the fields to explicitly demonstrate the QMB properties and stability. We have further argued that states with $q >2$ should also be stable. This is the first such example of stable $q>2$ monopoles away from the BPS limit that are bound nongravitationally. QMBs with large magnetic charge could exist with masses either well above or below the Planck scale. The theoretical structure required for such states to exist is surprisingly minimal. If a theory contains a magnetic monopole (already favored as an explanation for the quantization of electric charge), then QMB states can form given only one new gauge-singlet complex scalar field with a good $U(1)$ global symmetry and a renormalizable portal coupling to the gauged scalar of the monopole sector.

QMBs can form in a phase transition in the early universe. According to our estimates, QMBs formed in such a manner may have masses around or in excess of the Planck scale, as well as large radii far in excess of their intrinsic scale $1/v$. A mixture of QMBs and Q-balls may even make up all of dark matter. We have provided an analytic estimate for the abundance and average charge of QMBs following a phase transition. The existence of QMBs allows for new and interesting phenomena in their formation and evolution. First, Q-balls can aggregate monopoles as they cross through the interior of Q-balls and form possibly multiply magnetically charged bound states. Second, monopoles can act as seeds of Q-ball formation following a phase transition. Because the minimum stable global charge $Q_s$ is smaller for a QMB than for a Q-ball, sometimes as small as unity, it is easier to form QMBs from free $S$ particles and isolated monopoles than it is to form Q-balls from $S$ particle fusion \cite{Griest:1989bq}. Detailed numerical work similar to solitosynthesis on this formation mechanism, though beyond the scope of this work, would be of interest to verify our estimates. As an extension to this work, it would be interesting to examine the likelihood of forming magnetic black holes through this approach. If possible, the QMBs may serve as a novel formation mechanism for magnetic black holes in addition to the collapse of primordial density perturbations.

As the existence of monopole bound states could be a generic feature of high energy theories containing global charge, QMBs should be considered when devising and interpreting monopole searches. We have discussed how the Parker bound and direct monopole searches apply to QMBs. However, many more monopole search strategies exist \cite{Bai:2020spd,Mavromatos:2020gwk,Ghosh:2020tdu,Diamond:2021scl,Bai:2021ewf} or could be recast from similar searches \cite{VanDevender:2017avk,Sidhu:2019fgg,Sloan:2021pmf}. Several of these depend on the stopping power of monopoles in various materials (the Sun, the Earth, interstellar gases, etc.). 
QMBs may have electroweak or Grand Unified Theory symmetry restoration in their (potentially large radii) cores, giving them a potentially large geometric cross section with Standard Model matter or the ability to mediate baryon number violation. By establishing the existence of QMB states, we have shown that magnetically charged objects can have a larger variety of mass and magnetic charge than previously envisioned. Thus, an interesting topic of future study, beyond the scope of this work, would be to recast existing monopole bounds and to propose new search strategies for arbitrary values of mass, radius, and magnetic charge.

\subsubsection*{Acknowledgements}
The work of YB is supported by the U.S. Department of Energy under the contract DE-SC-0017647. The work of SL is supported in part by Israel Science Foundation under Grant No. 1302/19. The work of NO is supported by the Arthur B. McDonald Canadian Astroparticle Physics Research Institute.

%----------------------------------------------------------------
% References
%----------------------------------------------------------------
\setlength{\bibsep}{3pt}
\bibliographystyle{JHEP}
\bibliography{monopoleQball_refs}

\end{document}